\documentclass[12pt]{article}

\usepackage[utf8]{inputenc}
\usepackage[vcentermath]{youngtab}
\usepackage{color}
\usepackage{hyperref}
\usepackage[centertags]{amsmath}
\usepackage{amsfonts} 
\usepackage{amssymb}
\usepackage{amsthm}
\numberwithin{equation}{section} 
\usepackage{graphicx}
\usepackage[compat=1.1.0]{tikz-feynman}
\usepackage{contour}
\definecolor{light-gray}{gray}{0.95}
\usepackage{subcaption}

\begin{document}
\begin{titlepage}
\hfill \hbox{QMUL-PH-19-08}
\vskip 1.0cm
\begin{flushright}
\end{flushright}
\vskip 1.0cm
\begin{center}
  {\Large \bf 
    Revisiting the 2PM eikonal \\ and the dynamics of binary black holes}
   \vskip 1.0cm {\large Arnau Koemans Collado$^{a}$, Paolo Di Vecchia$^{b, c}$,
Rodolfo Russo$^{a}$} \\[0.7cm]
{\it $^a$ Centre for Research in String Theory, School of Physics and Astronomy \\ Queen Mary University of London, Mile End Road, E1 4NS London, United Kingdom}\\
{\it $^b$ The Niels Bohr Institute, University of Copenhagen, Blegdamsvej 17, \\
DK-2100 Copenhagen, Denmark}\\
{\it $^c$ Nordita, KTH Royal Institute of Technology and Stockholm University, \\Roslagstullsbacken 23, SE-10691 Stockholm, Sweden}\\
\end{center}

\begin{abstract}
In this paper we study the two-body gravitational scattering of massive scalars with different masses in general spacetime dimensions. We focus on the Regge limit (eikonal regime) of the resulting scattering amplitudes and discuss how to extract the classical information representing the scattering of two black holes. We derive the leading eikonal and explicitly show the resummation of the first leading energy contribution up to second order in Newton's gravitational constant. We also calculate the subleading eikonal showing that in general spacetime dimensions it receives a non-trivial contribution from the box integral. From the eikonal we extract the two-body classical scattering angle between the two black holes up to the second post-Minkowskian order (2PM). Taking various probe-limits of the two-body scattering angles we are able to show agreement between our results and various results in the literature. We highlight that the box integral also has a log-divergent (in energy) contribution at subsubleading order which violates perturbative unitarity in the ultra-relativistic limit. We expect this term to play a role in the calculation of the eikonal at the 3PM order.  
\end{abstract}

\end{titlepage}

\section{Introduction }
\label{sec:intro}

The high energy limit of scattering amplitudes in gravitational theories has been thoroughly studied as a gedanken-experiment that provides a non-trivial test of the consistency of the gravitational theory. A particularly tractable regime is the Regge limit, where both the energies and the impact parameter are large and unitarity is preserved due to a resummation of Feynman diagrams which reproduces the effect of a classical geometry~\cite{tHooft:1987vrq, Amati:1987wq,Muzinich:1987in,Sundborg:1988tb}. These early studies focused on the case of external massless states whose high energy Regge scattering matches the gravitational interaction of two well-separated shock-waves. However it is possible to generalise the same approach to the scattering of massive states~\cite{Kabat:1992tb} where the large centre of mass energy is due to both the kinetic and rest mass energy. It is then possible to interpolate between the ultra-relativistic Regge scattering mentioned above and the study of the non-relativistic large distance interaction between massive objects. This can be done both for pure General Relativity (GR) as well as for string theory, see for instance \cite{D'Appollonio:2010ae} for the analysis of the scattering of a perturbative massless state off a D-brane which we recall is a massive object\footnote{See~\cite{Bjerrum-Bohr:2014zsa,Bjerrum-Bohr:2016hpa} for the study of light/heavy scattering in standard GR including the derivation of quantum correction to the gravitational potential.}. The technique of deriving the relativistic interaction of two massive objects from an amplitude approach has recently attracted renewed attention~\cite{Neill:2013wsa,Akhoury:2013yua,Luna:2016idw,Cachazo:2017jef,Bjerrum-Bohr:2018xdl,Cheung:2018wkq,Kosower:2018adc,Bern:2019nnu} since it links directly to the post-Minkowskian approximation of the classical gravitational dynamics relevant for the inspiraling phase of binary black hole systems~\cite{Buonanno:1998gg,Damour:2016gwp,Damour:2017zjx,Antonelli:2019ytb}.

The amplitude approach to the relativistic two-body problem can be stated in the following conceptually simple way. Consider $2 \to 2$ scattering where the external states have the quantum numbers necessary to describe the classical objects one is interested in (massless states describe shock-waves, massive scalars can describe Schwarzschild black holes, then spin and charge can be added to describe Kerr\footnote{See~\cite{Guevara:2018wpp,Chung:2018kqs,Bautista:2019tdr} for a recent analysis of the amplitude approach to this case.} and Reissner-Nordstr\"om black holes). Then the limit is taken where the Newton's gravitational constant $G_N$ is small, but all classical parameters, such as the Schwarzschild radius or the classical angular momentum, are kept finite. Since in this paper we are interested in studying the scattering of scalar states, the only classical parameter in the problem is the effective Schwarzschild radius, $R_s^{D-3} \sim G_N M^*$, where $M^*$ is the largest mass scale in the process. We can have $M^*= \sqrt{s}$ in the ultra-relativistic/massless case or $M^*=m_1$ in the probe-limit with $m_1^2 \gg (s-m_1^2), m_2^2$. In either case the relevant kinematic regime is the Regge limit, since the centre of mass energy $\sqrt{s}$ has to be much larger than the momentum transferred $\sqrt{|t|}$. Since $G_N$ is small, one might think that the perturbative diagrams with graviton exchanges yield directly the effective two-body potential, but one must be careful in performing this step. In the limit mentioned above the perturbative amplitude at a fixed order in $G_N$ is divergent thus creating tension with unitarity. These divergent terms should exponentiate when resumming the leading contributions at large energy at different orders in $G_N$. This exponential, called the eikonal phase\footnote{In more general gravitation theories the eikonal phase can become an operator; this already happens~at~leading~order~in string theory~\cite{Amati:1987wq,Amati:1987uf,Amati:1988tn,D'Appollonio:2010ae} and also in an effective theory of gravity including higher derivative corrections \cite{Camanho:2014apa,DAppollonio:2015fly}.}, is the observable that we wish to calculate and that, as we will see, contains the relevant information for the two-body potential.

In this paper we will focus on the $2 \to 2$ scattering of massive scalar particles\cite{Kabat:1992tb,Akhoury:2013yua,Luna:2016idw,Bjerrum-Bohr:2018xdl} up to order $G_N^2$ (i.e. 2PM level). Here we keep the spacetime dimension $D$ general, which serves as an infrared regulator, and also consider the subleading ${\cal O}(G_N^2)$ contributions that do not directly enter in the 2PM classical interaction but that should be relevant for the 3PM result~\cite{Bern:2019nnu}. Since our analysis is $D$-dimensional we cannot apply the standard 4$D$ spinor-helicity description, but we construct the relevant parts of the amplitudes with one and two graviton exchanges by using an approach similar in spirit where tree-level amplitudes are glued together~\cite{Collado:2018isu,KoemansCollado:2019lnh}. The scaling limit discussed above can be spelled out for this case as follows:
\begin{itemize}
\item We take $G_N$ small by keeping $G_N M^*$ fixed and we are interested in the non-analytic contributions as $t\to 0$ since they determine the large distance interaction
\item The ratios $m_i^2/s$, where $m_{1,2}$ are the masses of the external scalars, can be arbitrary; when they are fixed, one is describing the scattering of two Schwarzschild black holes, but it is possible to smoothly take them to be small or large and make contact with different relativistic regimes
\item At each order in $G_N^n$ the terms that grow faster than $E_1$ or $E_2$ (at large $E_i$ and fixed $G_N M^*$) should not provide new data, but just exponentiate the energy divergent contributions at lower perturbative orders
\item The terms that grow as $E_i$ provide a new contribution to the eikonal phase at order $G_N^n$ from which one can derive the contribution to the classical two-body deflection angle and from it the relevant information on the $n$PM effective two-body potential
\end{itemize}
We carry out this approach explicitly up to the 2PM order. The $D$-dimensional case is slightly more intricate than the 4$D$ one as we find that the contribution from the scalar box integral not only contributes to the exponentiation of the 1PM result, but also yields non-trivial subleading terms that have to be combined with the triangle contributions to obtain the full 2PM eikonal. We also see that our result smoothly interpolates between the general, the light-bending (when $m_1^2 \gg s \gg m_2^2$) and  the ultra-relativistic cases (when $s\gg m^2_1,m^2_2$); this holds not just for the classical part of the 2PM eikonal phase, which is trivially zero in the massless case, but also for the quantum part~\cite{Amati:1990xe,Ciafaloni:2018uwe}. This feature does not seem to be realised in the recent 3PM result~\cite{Bern:2019nnu} and it would be interesting to understand this issue better.

\begin{figure}[h]
  \centering
  \begin{tikzpicture}[scale=1.5]
	\begin{feynman}
      \vertex[blob, minimum size=1.5cm] (m) at (0, 1) {\contour{white}{}};
      \vertex[blob, minimum size=1.5cm] (m2) at (0, -1) {\contour{white}{}};
	  \vertex[circle,inner sep=0pt,minimum size=0pt] (c) at (0, 0) {};      
      
      \vertex (a) at (2,1) {};
      \vertex (b) at (-2,1) {};
      \vertex (a2) at (2,-1) {};
      \vertex (b2) at (-2,-1) {};
	  \vertex (d) at (0.3,0.65) {};
	  \vertex (d2) at (-0.3,0.65) {};  
	  \vertex (e) at (0.3,-0.65) {};
	  \vertex (e2) at (-0.3,-0.65) {};      
      
      \diagram* {
      (a) -- [fermion,edge label=$k_3$, swap] (m),
      (b) -- [fermion,edge label=$k_1$] (m),
      (a2) -- [fermion,edge label=$k_4$, swap] (m2),
      (b2) -- [fermion,edge label=$k_2$] (m2),
      (d) -- [boson, half left, looseness=1.0] (e),
      (d2) -- [boson, half right, looseness=1.0] (e2),
      };
    \end{feynman}
    \end{tikzpicture}
  \caption{A figure illustrating the procedure outlined at the beginning of section \ref{sec:amplitude} and described by equation \eqref{eq:ampmaster} for the one-loop amplitude. The solid lines represent massive scalars and the wavy lines represent gravitons. The shaded blob is described by equation \eqref{eq:amp4pt}.}
  \label{fig:4ptglue}
\end{figure}
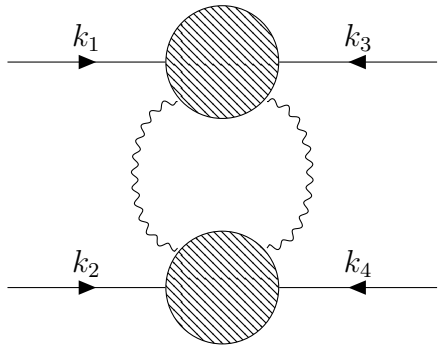

The paper is structured as follows. In section~\ref{sec:amplitude} we introduce the basic objects needed for our analysis, i.e. the tree-level on-shell vertices between two massive scalars and one and two gravitons. The field theory limit of a string expression provides a rather simple $D$-dimensional expression that we use to derive the relevant part of the amplitude with two graviton exchanges, see figure~\ref{fig:4ptglue}. We then extract the box and the triangle contributions that determine the 2PM eikonal phase. In section~\ref{sec:eikonal} we discuss the exponentiation pattern mentioned above and obtain explicit expressions for the 1PM and 2PM $D$-dimensional eikonal. As a check we derive the deflection angle in various probe-limits where it is possible to compare with a geodesic calculation in the metric of an appropriate black hole finding perfect agreement. Section~\ref{sec:discussion} contains a brief discussion on the possible relevance of our result for the study of the 3PM eikonal. In two appendices we provide the technical results needed in sections~\ref{sec:amplitude} and \ref{sec:eikonal}; in appendix~\ref{app:integrals} we evaluate the box and the triangle integrals in the limit  $s,m_i^2 \gg |t|$, while in appendix~\ref{sec:geodesics} we derive the deflection angle through a classical geodesic calculation in the background of a $D$-dimensional Schwarzschild black hole.

\section{Massive Scalar Scattering} \label{sec:amplitude}

In this section we focus on the $2 \rightarrow 2$ gravitational scattering process between two massive scalars in $D$ spacetime dimensions with both one and two graviton exchanges. As mentioned in the introduction, we are interested in extracting the classical contributions to this process, so instead of calculating the full amplitude by using the standard Feynman rules, we glue on-shell building blocks that capture just the unitarity cuts needed for reconstructing the classical eikonal. While this approach is by now commonly used in a $D=4$ setup, it is possible to implement it in general $D$ and here we follow~\cite{Collado:2018isu}, now including mass terms for the scalars.

For the one graviton exchange (1PM order) amplitude we can use, as an effective vertex, the on-shell three-point amplitude between two identical massive scalars and a graviton. In the standard Feynman vertex\footnote{We use the mostly plus convention for the metric and, as usual, we consider a two times signature to satisfy all on-shell constraints in the case of three-point functions.} $-i \kappa_D \left (k_{1 \mu}k_{2 \nu} + k_{1 \nu}k_{2 \mu} - (k_1 k_2 - m^2) \eta_{\mu \nu} \right )$, we can then drop the last two terms since they are proportional to $q^2$ and use the on-shell amplitude,
\begin{equation}
\begin{tikzpicture}[baseline=(eq)]
    \begin{feynman}[inline=(eq)]
      \vertex[blob, minimum size=1.0cm] (m) at ( 0, 0) {\contour{white}{}};
      \vertex (a) at (0,-2) {};
      \vertex (c) at (-2, 0) {};
      \vertex (d) at ( 2, 0) {};
      \vertex (eq) at (0,-1) {};
      \diagram* {
      (a) -- [boson,edge label=$q$] (m),
      (c) -- [fermion,edge label=$k_1$] (m),
      (d) -- [fermion,edge label=$k_2$, swap] (m),
      };
    \end{feynman}
\end{tikzpicture}
 = A_3^{\mu\nu}(k_1, k_2, q)
=  -i \kappa_D \left (k_{1}^{\mu}k_{2}^{\nu} + k_{1}^{\nu}k_{2}^{\mu} \right ) \;, \label{eq:amp3pt}
\end{equation}
where $\kappa_D = \sqrt{8 \pi G_N}$ and $G_N$ is the $D$-dimensional Newton's gravitational constant.

For two graviton exchange (2PM order) we need the corresponding four-point amplitude as the new ingredient. A particularly compact expression for this amplitude can be obtained by taking the field theory limit of the 2-tachyon 2-graviton amplitude in the Neveu-Schwarz string calculated by using the KLT approach. The result is,
\begin{eqnarray}
\hat{A}^{\alpha\beta;\rho\sigma}_4(k_1, k_2, q_1, q_2) = \frac{2\kappa_D^2 (k_2 q_1) (k_1 q_1)}{(q_1q_2) }\left[ \frac{k_2^\rho k_1^\alpha}{k_2 q_1}  + \frac{k_2^\alpha k_1^\rho}{k_1 q_1} +\eta^{\rho \alpha} \right]\left[ \frac{k_2^\sigma k_1^\beta}{k_2 q_1}  + \frac{k_2^\beta k_1^\alpha}{k_1 q_1} +\eta^{\sigma \beta} \right]\!.
\label{KLT24c}
\end{eqnarray}
By using the on-shell conditions it is possible to verify that~\eqref{eq:amp3pt} is symmetric under the exchange of the two scalars or the two gravitons and that it reproduces the known results for $D\to 4$, see for instance equations~(2.19) and~(2.2) of~\cite{Cachazo:2017jef}. For our purposes it will be convenient to use a different form for the amplitude where we have used momentum conservation and on-shell conditions to express $k_2$ in terms of $k_1$, $q_1$ and $q_2$,
\begin{eqnarray} \label{eq:amp4pt}
&&
\begin{tikzpicture}[baseline=(eq)]
    \begin{feynman}[inline=(eq)]
      \vertex[blob, minimum size=1.0cm] (m) at ( 0, 0) {\contour{white}{}};
      \vertex (a) at (-1,-2) {};
      \vertex (b) at ( 1,-2) {};
      \vertex (c) at (-2, 0) {};
      \vertex (d) at ( 2, 0) {};
      \vertex (eq) at (0,-1) {};
      \diagram* {
      (a) -- [boson,edge label=$q_1$] (m),
      (b) -- [boson,edge label=$q_2$,swap] (m),
      (c) -- [fermion,edge label=$k_1$] (m),
      (d) -- [fermion,edge label=$k_2$, swap] (m),
      };
    \end{feynman}
\end{tikzpicture}
 = A_4^{\alpha\beta;\rho\sigma}(k_1, k_2, q_1, q_2) \nonumber \\
 &&=\frac{-2\kappa_D^2 [(k_1 q_1) + (q_2 q_1)] (k_1 q_1)}{(q_1q_2) } \biggl( \frac{(k_1 + q_2 )^\rho k_1^\alpha}{(k_1 q_1) + (q_2 q_1)}  - \frac{(k_1  +q_1)^\alpha k_1^\rho}{k_1 q_1} +\eta^{\rho \alpha} \biggr) \nonumber \\
&& \qquad \times \biggl( \frac{(k_1 + q_2 )^\sigma k_1^\beta}{(k_1 q_1) + (q_1 q_2)}  - \frac{(k_1 +q_1)^\beta k_1^\alpha}{k_1 q_1} +\eta^{\sigma \beta} \biggr) \;.
\end{eqnarray}
Of course this expression is equivalent to~\eqref{KLT24c} on-shell, but~\eqref{eq:amp4pt} is transverse in the following slightly more general sense: it vanishes whenever the polarization of a graviton takes the form $\epsilon_{\mu\nu}=\zeta_{\mu} q_\nu+\zeta_\nu q_\mu$ just by using the on-shell conditions and  momentum conservation to rewrite products between momenta such as $k_i k_j$ (without the need of using it to rewrite the products between momenta and the arbitrary vectors $\zeta_i$).

In the next subsection we derive the classical ${\cal O}(G_N)$ contribution by gluing two amplitudes~\eqref{eq:amp3pt} with the de Donder propagator,
\begin{equation}\label{eq:dedpro}
[G(q)]^{\mu \nu; \rho \sigma} = \frac{-i}{2q^2} \left ( \eta^{\mu \rho} \eta^{\nu \sigma} + \eta^{\mu \sigma} \eta^{\nu \rho} - \frac{2}{D-2} \eta^{\mu \nu} \eta^{\rho \sigma} \right )\,.
\end{equation} 
In subsection \ref{sec:oneloopamp} we obtain the ${\cal O}(G_N^2)$ result by gluing the gravitons of two copies of the amplitude~\eqref{eq:amp4pt}. In all the four scalar amplitudes obtained in this section we denote the two incoming particles with momenta $k_1$ and $k_2$ and outgoing momenta $k_3$ and $k_4$. The particles $1$ and $3$ have mass $m_1$, while the particles $2$ and $4$ have mass $m_2$, see for instance figure~\ref{fig:3ptglue}. Finally we will use the following Mandelstam variables throughout this paper,
\begin{equation}
s=-(k_1+k_2)^2 ~~;~~ u=-(k_1+k_4)^2 ~~;~~ t=-(k_1+k_3)^2 \;.
\end{equation}

\subsection{One Graviton Exchange} \label{sec:treelevelamp}

Using the gluing procedure outlined above we can calculate the tree-level four-point massive scalar scattering by gluing two amplitudes~\eqref{eq:amp3pt} with a de Donder propagator~\eqref{eq:dedpro} and obtain,
\begin{equation}\label{eq:amptree}
i \mathcal{A}_1 = [G(k_1+k_3)]_{\mu_1 \nu_1; \mu_2 \nu_2}\; A^{\mu_1\nu_1}_3 (k_1, k_3, -k_1-k_3) A_3^{\mu_2\nu_2}(k_2, k_4, k_1+ k_3) \;.
\end{equation}
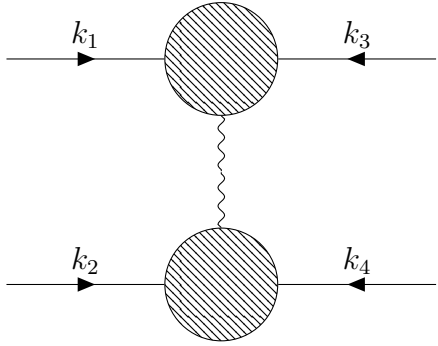
\begin{figure}[h]
  \centering
  \begin{tikzpicture}[scale=1.5]
	\begin{feynman}
      \vertex[blob, minimum size=1.5cm] (m) at (0, 1) {\contour{white}{}};
      \vertex[blob, minimum size=1.5cm] (m2) at (0, -1) {\contour{white}{}};
	  \vertex[circle,inner sep=0pt,minimum size=0pt] (c) at (0, 0) {};      
      
      \vertex (a) at (2,1) {};
      \vertex (b) at (-2,1) {};
      \vertex (a2) at (2,-1) {};
      \vertex (b2) at (-2,-1) {};
      \diagram* {
      (a) -- [fermion,edge label=$k_3$, swap] (m),
      (b) -- [fermion,edge label=$k_1$] (m),
      (a2) -- [fermion,edge label=$k_4$, swap] (m2),
      (b2) -- [fermion,edge label=$k_2$] (m2),
      (c) -- [boson] (m),
      (c) -- [boson] (m2),
      };
    \end{feynman}
    \end{tikzpicture}
  \caption{A figure illustrating the procedure outlined at the beginning of section \ref{sec:amplitude} and described by equation \eqref{eq:amptree} for the tree-level amplitude. The solid lines represent massive scalars and the wavy lines represent gravitons. The shaded blob is described by equation \eqref{eq:amp3pt}.}
  \label{fig:3ptglue}
\end{figure}
We then find,
\begin{equation}
i \mathcal{A}_{1} = \frac{2i \kappa_D^2}{q^2} \left ( \frac{1}{2}(s- m_1^2 - m_2^2)^2 - \frac{2}{D-2} m_1^2 m_2^2 \right )  =  \frac{2 i \kappa_D^2 \gamma(s)}{q^2} \label{eq:1geamp} \;,
\end{equation}
where $q \equiv k_1+k_3$ is the momentum exchanged between the two massive scalars and we have defined the quantity,
\begin{equation}
\gamma(s)=2(k_1 k_2)^2 - \frac{2}{D-2} m_1^2 m_2^2 = \frac{1}{2} (s-m_1^2-m_2^2)^2 - \frac{2}{D-2} m_1^2 m_2^2 \;. \label{eq:1gegms}
\end{equation}
In the high energy limit and after moving into impact parameter space (as defined below) we can see that this contribution grows as $E_i$ (since $G_N M^*$ is constant) and violates perturbative unitarity at large energies, we will come back to this point when discussing the two graviton exchange amplitude as well as in section~\ref{sec:eikonal}. By construction, this result just captures the pole contribution in $t$ of the amplitude, but this is sufficient to extract the classical interaction between two well separated particles. This is more clearly seen by transforming the amplitude to impact parameter space. As is standard in the discussion of the eikonal phase, we introduce an auxiliary $(D-2)$-dimensional vector $\mathbf{q}$ such that $\mathbf{q}^2=-t$ and then take the Fourier transform to rewrite the result in terms of the conjugate variable $\mathbf{b}$ (the impact parameter). We can then calculate the amplitude in impact parameter space by using,
\begin{equation}
\tilde{\mathcal{A}} = \frac{1}{4 E p} \int \frac{{\textrm{d}}^{D-2}\mathbf{q}}{(2\pi)^{D-2}}e^{i\mathbf{q} \mathbf{b}} \mathcal{A} \label{eq:ampips}\;,
\end{equation}
where $E=E_1+E_2$ and $p=|p_1|=|p_2|$ is the absolute value of the space-like momentum in the center of mass frame of the two scattering particles. We can therefore also calculate,
\begin{equation}\label{eq:1geEP}
2 E p = 2 \sqrt{(k_1 k_2)^2 - k_1^2 k_2^2} = \sqrt{(s-m_1^2 -m_2^2)^2 - 4 m_1^2 m_2^2} \;.
\end{equation}
Terms in~\eqref{eq:1geamp} that are regular as we take $t\to 0$ yield only delta-function contributions localised at $\mathbf{b}=0$ and so can be neglected. Here and in the following sections the following integral will be useful when computing impact parameter space expressions,
\begin{eqnarray}
\int \frac{d^d \mathbf{q}}{(2\pi)^d} {\rm e}^{i  \mathbf{q}  \mathbf{b}} ( \mathbf{q}^2)^\nu = \frac{2^{2\nu}}{\pi^{d/2}}
\frac{\Gamma ( \nu + \frac{d}{2})}{\Gamma (-\nu)} 
\frac{1}{ ( \mathbf{b}^2)^{\nu + \frac{d}{2}}} \;. \label{eq:impoformu}
\end{eqnarray}
We can now use \eqref{eq:ampips} and \eqref{eq:impoformu} to find the impact parameter space expression of the tree-level contribution,
\begin{equation}
i \tilde{\mathcal{A}}_{1} = \frac{i \kappa_D^2 \gamma(s)}{2 E p} \frac{1}{4 \pi^{\frac{D-2}{2}}} \Gamma \left( \frac{D}{2}-2 \right) \frac{1}{ \mathbf{b}^{D-4}} \;. \label{eq:1geimp}
\end{equation}
This result agrees with known results \cite{Kabat:1992tb} and as discussed in more detail in section \ref{sec:eikonal} is related to the result for the first order contribution to the deflection angle in the post-Minkowskian expansion.

\subsection{Two Graviton Exchanges} \label{sec:oneloopamp}

In this subsection we discuss the gluing procedure at one-loop. Schematically we have,
\begin{eqnarray}\label{eq:ampmaster}
i \mathcal{A}_2 &=& \int \frac{d^{D} k}{(2\pi)^{D}} [G(k)]_{\alpha_1 \beta_1 ; \alpha_2 \beta_2}\; [G(k+q)]_{\rho_1 \sigma_1 ; \rho_2 \sigma_2} \; \nonumber \\
&& \qquad \qquad \times A_4^{\alpha_1 \beta_1 ; \rho_1 \sigma_1} (k_1, k_3, k, -k-q) A_4^{\alpha_2 \beta_2 ; \rho_2 \sigma_2}(k_2, k_4, -k, k+q) \;.
\end{eqnarray}
where $A_4$ is the four-point amplitude given by \eqref{eq:amp4pt}, we recall that $q \equiv k_1+k_3$ is the momentum exchanged between the two massive scalars, $k$ is the momentum in the loop and $[G]$ represents the graviton propagator~\eqref{eq:dedpro}.

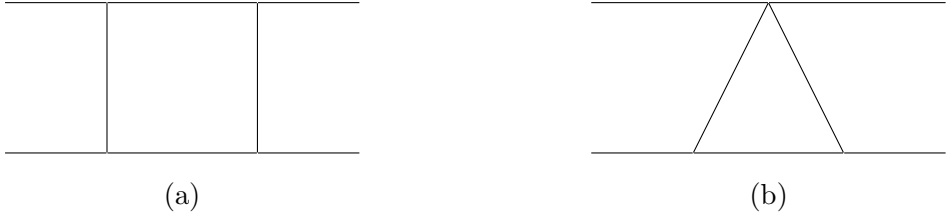
\begin{figure}[h]
  \begin{subfigure}[t]{0.45\textwidth}
    \centering
      \begin{tikzpicture}
	    \begin{feynman}
			\vertex (a) at (-2.5,-2) {};
			\vertex (b) at ( 2.5,-2) {};
			\vertex (c) at (-2.5, 0) {};
			\vertex (d) at ( 2.5, 0) {};
			\vertex[circle,inner sep=0pt,minimum size=0pt] (e) at (-1, 0) {};
			\vertex[circle,inner sep=0pt,minimum size=0pt] (f) at (1, 0) {};
			\vertex[circle,inner sep=0pt,minimum size=0pt] (g) at (-1, -2) {};
			\vertex[circle,inner sep=0pt,minimum size=0pt] (h) at (1, -2) {}; 
			\diagram* {
			(c) -- (e) -- (f) -- (d),
			(g) -- (e),
			(h) -- (f),
			(a) -- (g) -- (h) -- (b),
			};
	    \end{feynman}
  \end{tikzpicture}
    \caption{}
    \label{fig:amp1}
  \end{subfigure}
  \quad
  \begin{subfigure}[t]{0.45\textwidth}
    \centering
    \begin{tikzpicture}
		\begin{feynman}
			\vertex (c) at (-2.5, 0) {};
			\vertex (d) at ( 2.5, 0) {};
			\vertex (x) at (-2.5, -2) {};
			\vertex (y) at ( 2.5, -2) {};
			\vertex[circle,inner sep=0pt,minimum size=0pt] (e) at (0, -1) {};
			\vertex[circle,inner sep=0pt,minimum size=0pt] (m) at (0, 0) {};
			\vertex[circle,inner sep=0pt,minimum size=0pt] (a) at (-1, -2) {};
			\vertex[circle,inner sep=0pt,minimum size=0pt] (b) at (1, -2) {}; 
     
			\diagram* {
			(c) -- (m) -- (d),
			(a) -- (m),
			(b) -- (m),
			(x) -- (a) -- (b) -- (y),
			};
		\end{feynman}
  \end{tikzpicture}
    \caption{}
    \label{fig:amp2}
  \end{subfigure}
  \caption{The two topologies of integrals that contribute to the two graviton exchange amplitude in the classical limit. In \ref{fig:amp1} we have the box topology and in \ref{fig:amp2} we have the triangle topology. The integral structure in \ref{fig:amp2} receives contributions from various Feynman diagrams, including those with a three-point vertex in the bulk. We can ignore other integral structures, such as bubble and tadpoles, since they do not contribute in the classical limit.}
  \label{fig:1}
\end{figure}

In order to interpret the expression found after attaching the relevant vertices using \eqref{eq:ampmaster} we need to rewrite it in terms of the relevant integral topologies which are schematically shown in figure \ref{fig:1}. In order to do this we define an operation denoted as $\mathcal{S}_{n}[\mathcal{A}_2]$ which searches the full expression, $\mathcal{A}_2$ resulting from \eqref{eq:ampmaster} and yields the integrand with $n$ number of propagators. Starting from the maximum number of propagators which in this case is $n=4$, we have,
\begin{equation}
\mathcal{S}_{4}[\mathcal{A}_2] = a_{\square} = \int \frac{d^{D} k}{(2\pi)^{D}} \frac{1}{k^2} \frac{1}{(q+k)^2} \frac{1}{(k_1 + k)^2 + m_1^2} \frac{1}{(k_2 - k)^2 + m_2^2} \mathcal{N}_{\square} \;,
\end{equation}
where we have set all the momenta in the internal propagators on-shell in $\mathcal{N}_{\square}$ since terms proportional to any propagator would cancel with one of the propagators in the denominator and therefore not contribute to the diagram with the above pole structure. Note that we have identified the pole structure above with the so called scalar box integral topology. An explicit expression for the numerators will be given in the upcoming subsections. 

We now want to search further in order to find the integrand with 3 poles. So now we have,
\begin{equation}
\mathcal{S}_{3}[\mathcal{A}_2 - a_{\square}] = a_{\triangle} = \int \frac{d^{D} k}{(2\pi)^{D}} \frac{1}{k^2} \frac{1}{(q+k)^2} \frac{1}{(k_1 + k)^2 + m_1^2} \mathcal{N}_{\triangle} \;,
\end{equation}
where we are searching the difference between the full expression, $\mathcal{A}_2$, and the part already extracted for the box diagram, $a_{\square}$. We have also set the momenta in the internal propagators on-shell in, $\mathcal{N}_{\triangle}$, for the same reasons described previously. Note that we have identified the pole structure above with the so called triangle integrals. It should be mentioned that one also extracts the crossed box and "inverted" triangle (i.e. the contribution with the opposite massive scalar propagator) by searching for the relevant pole structures.

Once the procedure described above has been completed the classical contributions to each of the expressions above are determined by implementing the scaling limit mentioned in the introduction,
\begin{equation}
  \label{eq:heml}
  16 \pi G_N=2\kappa^2 \to 0~,~~~s\gg q^2=|t|~,~~~\mbox{with}~\,G_N M^*~\,\mbox{fixed}\;.
\end{equation}
For two graviton exchanges we have two amplitude topologies that contribute; the box and triangle integrals, which are shown in figure \ref{fig:1}. The masses can be of the same order or much smaller than the centre of mass energy and of course the integrals take different forms in these two cases. In appendix \ref{app:integrals} we focus on the case $s\sim m_i^2$ and evaluate the first terms in the high energy expansion~\eqref{eq:heml} for the box and triangle integrals. In the ultra-relativistic case one recovers the massless results that can be found for instance in~\cite{Ellis:2007qk}.

\subsubsection{Box contribution}\label{sec:boxamplitude}

From the procedure outlined at the start of this subsection we find the following expression for the numerator, $\mathcal{N}_{\square}$, of the box diagram contribution to the two graviton exchange amplitude,
\begin{equation}
\mathcal{N}_{\square} = 4 \kappa_D^4 \gamma^2(s) \;,
\end{equation}
where $\gamma(s)$ has been defined in \eqref{eq:1gegms}. Writing this by including the integration over the loop momenta as well as including the contribution from the crossed box diagram we find,~
\begin{equation}\label{eq:fullboxamp}
i \mathcal{A}_{2} = 4 \kappa_D^4 (\gamma^2(s) \mathcal{I}_4(s,t) + \gamma^2(u) \mathcal{I}_4(u,t)) \;.
\end{equation}
where the integrals $\mathcal{I}_4(s,t)$ and $\mathcal{I}_4(u,t)$ have been computed in detail in appendix \ref{app:boxintegrals} and are defined as,
\begin{eqnarray}
\mathcal{I}_4(s,t) = \int \frac{d^{D} k}{(2\pi)^{D}} \frac{1}{k^2} \frac{1}{(q+k)^2} \frac{1}{(k_1 + k)^2 + m_1^2} \frac{1}{(k_2 - k)^2 + m_2^2} \;, \\
\mathcal{I}_4(u,t) = \int \frac{d^{D} k}{(2\pi)^{D}} \frac{1}{k^2} \frac{1}{(q+k)^2} \frac{1}{(k_3 + k)^2 + m_1^2} \frac{1}{(k_2 - k)^2 + m_2^2} \;.
\end{eqnarray}
Substituting the results for the integrals we find the leading contribution in the limit described by \eqref{eq:heml},
\begin{equation}
i \mathcal{A}^{(1)}_{2} =  -\frac{\pi^{\frac{D}{2}}}{(2\pi)^D} \frac{\pi}{2} \frac{4 \kappa_D^4 \gamma^2(s)}{\sqrt{(k_1 k_2)^2 - m_1^2 m_2^2}} \frac{\Gamma^2(\frac{D}{2}-2) \Gamma(3-\frac{D}{2})}{\Gamma(D-4)} (q^2)^{\frac{D}{2}-3} \;.
\end{equation}
The details of how to take the limit described by \eqref{eq:heml} when performing the integrals required to yield this result is given in appendix \ref{app:boxintegrals}. Moving to impact parameter space using \eqref{eq:ampips} and \eqref{eq:impoformu} we find that,
\begin{equation}
i \tilde{\mathcal{A}}^{(1)}_{2} = - \frac{\kappa_D^4 \gamma^2(s)}{(Ep)^2} \frac{1}{2^7 \pi^{D-2}} \Gamma^2 \left(\frac{D}{2}-2 \right) \frac{1}{\mathbf{b}^{2D-8}} \;.
\label{eq:boxleading}
\end{equation}
In the limit~\eqref{eq:heml}, this contribution grows as $E_i^2$ (since $G_N M^*$ is constant). Comparing with \eqref{eq:1geimp} we easily see that $i \tilde{\mathcal{A}}^{(1)}_{2}=\frac{1}{2} (i \tilde{\mathcal{A}}_{1})^2$, which is the first sign of the eikonal exponentiation discussed in more detail in section \ref{sec:eikonal}; the exponential of the tree-level amplitude will account for the first leading energy contributions of all higher loop amplitudes.

We can also look at the subleading contribution, in the limit described by \eqref{eq:heml}, to the two graviton exchange box diagram (as we will see in section \ref{sec:eikonal} this contributes to the second order of the post-Minkowskian expansion). Using the result for the subleading contribution to the integrals $\mathcal{I}_4(s,t)$ and $\mathcal{I}_4(u,t)$ found in  \eqref{eq:ij4subleading} we have,
\begin{equation}\label{eq:boxsubleadingbq}
i \mathcal{A}^{(2)}_{2} = \frac{i 2 \kappa_D^4 \gamma^2(s) \sqrt{\pi}}{(4\pi)^{\frac{D}{2}}} \frac{m_1+m_2}{(k_1 k_2)^2 - m_1^2 m_2^2} \frac{\Gamma \left(\frac{5-D}{2}\right) \Gamma^2\left(\frac{D-3}{2}\right)}{\Gamma(D-4)} (q^2)^{\frac{D-5}{2}}\;.
\end{equation}
At large energies this result scales as $E_i$ exactly as $\mathcal{A}_{1}$. This contribution should be exponentiated by the first subleading terms in the energy expansion of the higher loop contributions and so provides a new contribution to the eikonal phase. In impact parameter space~\eqref{eq:boxsubleadingbq} becomes,
\begin{equation}
i \tilde{\mathcal{A}}^{(2)}_{2} = \frac{i \kappa_D^4 \gamma^2(s)(m_1+m_2)}{Ep((k_1 k_2)^2 - m_1^2 m_2^2)} \frac{1}{64 \pi^{D-\frac{3}{2}}} \frac{\Gamma \left(\frac{2D-7}{2}\right) \Gamma^2\left(\frac{D-3}{2}\right)}{\Gamma(D-4)} \frac{1}{\mathbf{b}^{2D-7}} \;. \label{eq:boxsubleadingb}
\end{equation}
We have checked that the results in this subsection agree in $D=4$ with equivalent results \cite{Kabat:1992tb, Akhoury:2013yua, Luna:2016idw,Damour:2017zjx,Bjerrum-Bohr:2018xdl}. Let us stress that equation~\eqref{eq:boxsubleadingb} vanishes in the $D\to 4$ limit because of the presence of the factor of $\Gamma(D-4)$ in the denominator. Thus, for $D>4$ there is a contribution to the eikonal from the box integral which becomes trivial in the four dimensional case. In general, this contribution is crucial in order to match, in the probe-limit, with the geodesic calculations as discussed in section~\ref{sec:eikonal} and in \cite{Collado:2018isu} for the massless case $m_1 \gg 0, m_2=0$.

The subsubleading contributions to the box diagram are naively expected to be finite in the limit described by \eqref{eq:heml}, but there is actually a log-divergent term in the amplitude, as discussed for the massless case in~\cite{Amati:1990xe,Ciafaloni:2018uwe}, see also~\cite{Bellini:1992eb,Dunbar:1994bn} for an explicit evaluation of the same $2\to 2$ one-loop process with external gravitons. This contribution comes from using \eqref{eq:fullboxamp} and the next order in the expansion of the box integral, which in our case yields,
\begin{eqnarray}\label{eq:boxsubsubleadingbq}
&& i \mathcal{A}^{(3)}_{2} = 4 \kappa_D^4 \gamma^2 (s) \frac{i}{8(4\pi)^{\frac{D}{2}}} \Gamma \left(\frac{4-D}{2}\right)\, \frac{\Gamma^2\!\left(\frac{D-2}{2}\right)}{\Gamma(D-4)}  (q^2)^{\frac{D-4}{2}} \frac{1}{D-4} \nonumber \\
&& \times \left[ \frac{4(5-D)}{(k_1 k_2)^2 - m_1^2 m_2^2} \left( 1 + \frac{2k_1 k_2 \, \text{arcsinh} \left( \sqrt{\frac{\sigma-1}{2}} \right)}{\sqrt{(k_1k_2)^2 - m_1^2 m_2^2}}  \right)  + i \frac{\pi (D-4) (k_1+k_2)^2}{[(k_1k_2)^2 - m_1^2 m_2^2]^{3/2}} \right] \nonumber \\
&& - 4 \kappa_D^4 \psi(s)  \frac{i}{(4\pi)^{\frac{D}{2}}} \frac{\text{arcsinh}\left(\sqrt{\frac{\sigma-1}{2}}\right)}{ \sqrt{(k_1k_2)^2 - m_1^2 m_2^2}}\,\Gamma \left(\frac{6-D}{2}\right)\, \frac{\Gamma^2\!\left(\frac{D-4}{2}\right)}{\Gamma(D-4)} (q^2)^{\frac{D-4}{2}} \;,
\end{eqnarray}
where we have defined $\sigma = \frac{- k_1 k_2}{m_1 m_2}$ and,
\begin{eqnarray}
\psi(s) &=& -(2 k_1 k_2)\left( (2 k_1 k_2)^2 -\frac{4 m_1^2 m_2^2}{D-2} \right) \nonumber \\
&=& \left(s-m_1^2-m_2^2\right) \left( \left(s-m_1^2-m_2^2\right)^2 -\frac{4 m_1^2 m_2^2}{D-2} \right) \;.
\end{eqnarray}
Note that the last term in \eqref{eq:boxsubsubleadingbq} comes from expressing the $\gamma^2(u)$ from the second term in \eqref{eq:fullboxamp} in terms of $\gamma^2(s)$, i.e. we have $\gamma^2(u)=\gamma^2(s)+t\,\psi(s)+\mathcal{O}(t^2)$. We can also write the result above in impact parameter space for which we find,
\begin{eqnarray}\label{eq:boxsubsubleadingbqips}
&& i \tilde{\mathcal{A}}^{(3)}_{2} = \frac{\kappa_D^4 \gamma^2 (s)}{Ep} \frac{i}{128 \pi^{D-1}} \Gamma^2\!\left(\frac{D-2}{2}\right) \frac{1}{(\mathbf{b}^2)^{D-3}} \nonumber \\
&& \times \left[ \frac{4(5-D)}{(k_1 k_2)^2 - m_1^2 m_2^2} \left( 1 + \frac{2k_1 k_2 \, \text{arcsinh} \left( \sqrt{\frac{\sigma-1}{2}} \right)}{\sqrt{(k_1 k_2)^2 - m_1^2 m_2^2}}  \right) + i \frac{\pi (D-4) (k_1+k_2)^2}{[(k_1k_2)^2 - m_1^2 m_2^2]^{3/2}} \right] \nonumber \\
&& + \frac{\kappa_D^4 \psi(s)}{Ep} \frac{i}{8 \pi^{D-1}} \frac{\text{arcsinh}\left(\sqrt{\frac{\sigma-1}{2}}\right)}{ \sqrt{(k_1 k_2)^2 - m_1^2 m_2^2}} \Gamma^2\!\left(\frac{D-2}{2}\right) \frac{1}{(\mathbf{b}^2)^{D-3}} \;.
\end{eqnarray}
By using $\text{arcsinh} \, y=\log(y+\sqrt{y^2+1})$ in equation \eqref{eq:boxsubsubleadingbq} we can see that the second term on the second line and the term on the last line are log-divergent at large energies. It is interesting to highlight the following points. First, the same arcsinh-function arising from this subsubleading contribution also appears in the recent 3PM result~\cite{Bern:2019nnu}. Then these terms violate perturbative unitarity in the $s/m_i^2 \to \infty$ limit and~\cite{Amati:1990xe} conjectured that they should resum to provide a quantum correction to the eikonal phase. This contribution is relevant in the discussion of the Reggeization of the graviton, for a recent discussion see~\cite{Melville:2013qca} and references therein. Finally the contribution~\ref{eq:boxsubsubleadingbqips} provides an additional imaginary part to $\tilde{\cal A}_2$ beside that coming from the leading term~\eqref{eq:boxleading}. In~\cite{Amati:1990xe}, it was shown that this subleading imaginary part vanishes in the $D=4$ massless case. Since the last term in the second line vanishes in $D=4$, here we find through a direct calculation that the same result holds also for the scattering of massive scalars. We will briefly come back to these points in section~\ref{sec:discussion}.

\subsubsection{Triangle contribution}
\label{sec:triangleamplitude}

Following the procedure outlined at the beginning of this subsection we find that the expression for the numerator, $\mathcal{N}_{\triangle}$, for the triangle-like contributions, with the $m_1$ massive scalar propagator, is given by,
\begin{eqnarray}
\mathcal{N}_{\triangle} &=& \kappa_D^4 \left( \frac{16(D-3) (k \, k_2)^2 m_1^4}{(D-2)q^2} \right. \nonumber \\ 
&& \quad \left. + 4m_1^2 \left[ 2m_1^2 m_2^2 \frac{D^2 - 4D + 2}{(D-2)^2} - 2m_1^2 s  + m_1^4 + (m_2^2-s)^2 \right] \right) \;, \nonumber \\
\label{eq:triintgrand}
\end{eqnarray}
where we have already neglected some terms which are subleading in the limit given by \eqref{eq:heml} (i.e. don't contribute classically at second post-Minkowskian order). As we've done before we can express this in terms of an integral basis in which the expression \eqref{eq:triintgrand} becomes,
\begin{eqnarray}
&&\kappa_D^4 \biggl\{ \frac{16(D-3) k_{2 \mu} k_{2 \nu} m_1^4}{(D-2)q^2} \mathcal{I}_{3}^{\mu \nu}(m_1) \nonumber \\
&& + 4m_1^2 \left[ 2m_1^2 m_2^2 \frac{D^2 - 4D + 2}{(D-2)^2} - 2m_1^2 s  + m_1^4 + (m_2^2-s)^2 \right] \mathcal{I}_{3}(m_1) \biggr\} \nonumber \\
&& \qquad + m_1 \leftrightarrow m_2 \;,
\end{eqnarray}
where we recall that $k$ is the loop momentum and we have now included the contribution coming from the equivalent diagram with the $m_2$ massive scalar propagator. We have also defined the integrals,
\begin{eqnarray}
\mathcal{I}_{3}^{\mu \nu}(m_i) &=& \int \!\frac{d^Dk}{(2\pi)^D}\, \frac{1}{k^2}  \,\frac{1}{(q+k)^2} \, \frac{1}{(k+k_i)^2 + m_i^2} k^{\mu} k^{\nu} \;, \\
\mathcal{I}_{3}(m_i) &=& \int \!\frac{d^Dk}{(2\pi)^D}\, \frac{1}{k^2} \,\frac{1}{(q+k)^2} \,\frac{1}{(k+k_i)^2+m_i^2} \;.
\end{eqnarray}
Substituting the appropriate results for these integrals in the limit described by \eqref{eq:heml}, which are calculated in appendix \ref{app:triintegrals}, yields,
\begin{eqnarray}
i\mathcal{A}^{(2)}_2 &=& i \frac{2 \kappa_D^4 \sqrt{\pi}}{(4\pi)^{\frac{D}{2}}} \frac{\Gamma{\left( \frac{5-D}{2} \right)} \Gamma^2{\left( \frac{D-3}{2} \right)}}{\Gamma{\left( D-3 \right)}}(q^2)^{\frac{D-5}{2}} (m_1+m_2) \biggl\{  (s-m_1^2 -m_2^2)^2 \nonumber \\
&& - \frac{4m_1^2 m_2^2 }{(D-2)^2}  - \frac{(D-3) \left( (s-m_1^2 -m_2^2)^2 - 4 m_1^2 m_2^2 \right)  }{4 (D-2)^2} \biggr\} \label{eq:triamp} \;,
\end{eqnarray}
where we have again neglected subleading terms which do not contribute at second post-Minkowskian order. We can write equation \eqref{eq:triamp} in impact parameter space,
\begin{eqnarray}
i \tilde{\mathcal{A}}^{(2)}_{2} &=& i \frac{\kappa_D^4}{64 \pi^{D-\frac{3}{2}} \, Ep} \frac{\Gamma \left(\frac{2D-7}{2}\right) \Gamma^2\left(\frac{D-3}{2}\right)}{\Gamma(D-3)} \frac{m_1+m_2}{\mathbf{b}^{2D-7}} \biggl\{ (s-m_1^2 -m_2^2)^2 \nonumber \\
&& - \frac{4m_1^2 m_2^2 }{(D-2)^2}  - \frac{(D-3) \left( (s-m_1^2 -m_2^2)^2 - 4 m_1^2 m_2^2 \right)  }{4 (D-2)^2} \biggr\} \;. \label{eq:triips}
\end{eqnarray}
The results in this subsection agree with results for $D=4$ found in \cite{Luna:2016idw,Damour:2017zjx,Bjerrum-Bohr:2018xdl}. We have not considered the subleading triangle-like contribution explicitly in this subsection because we have found that it does not contribute to the log-divergent terms we discuss in sections \ref{sec:boxamplitude} and \ref{sec:discussion}. This should be clear from the results for the various integrals in appendix \ref{app:triintegrals}. Note also that these subleading contributions do not produce contributions to the real part of $i \mathcal{A}_2$.

\section{The Eikonal and Two-Body Deflection Angles} \label{sec:eikonal}

In this section we summarise general expressions for the eikonal and the deflection angle for the case of two different masses.  We then discuss explicit expressions for these quantities using the amplitudes derived in section \ref{sec:amplitude}. We also discuss various probe-limits for both general $D$ and $D=4$ in order to compare with existing results in the literature.

We will start by defining what the eikonal phase is in the context we are considering in this paper. We recall from section \ref{sec:amplitude} that the amplitudes in impact parameter space are defined via,
\begin{equation}
\tilde{\mathcal{A}}_{n}(s,m_i,\mathbf{b}) = \frac{1}{4 E p} \int \frac{{\textrm{d}}^{D-2}\mathbf{q}}{(2\pi)^{D-2}}e^{i\mathbf{q} \mathbf{b}}  \mathcal{A}_{n}(s,m_i,q) \label{eq:eikips}\;,
\end{equation}
where the various symbols have been defined previously and we have that $\mathcal{A}_n$ is the appropriate amplitude with $n$ graviton exchanges. We can generally write the gravitational S-matrix in impact parameter space as \cite{Giddings:2009gj,Giddings:2011xs},
\begin{equation}
S(s,m_i,\mathbf{b}) = 1 + i \sum_{n=1}^{\infty} \tilde{\mathcal{A}}_{n}(s,m_i,\mathbf{b})
\end{equation} 
where $\tilde{\mathcal{A}}_{n}(s,m_i,\mathbf{b})$ is the full amplitude with $n$ graviton exchanges in impact parameter space including the appropriate normalisation as defined in \eqref{eq:eikips}. As was discussed in \cite{Collado:2018isu} the gravitational S-matrix in the eikonal approximation can be expressed as,
\begin{eqnarray}
S(s,m_i,\mathbf{b})  &=&  \left(1 + i T(s,m_i,\mathbf{b}) \right) \exp \left[{i (\delta^{(1)}(s,m_i,\mathbf{b}) +  \delta^{(2)}(s,m_i,\mathbf{b}) + \ldots)}\right] \text{ ,} \label{eq:eiksmatrix} 
\end{eqnarray}
where $\delta^{(1)}(s,m_i,\mathbf{b})$ and $\delta^{(2)}(s,m_i,\mathbf{b})$ are the leading eikonal and subleading eikonal respectively. The parameter used to define the expansion in~\eqref{eq:eiksmatrix} is $(R_s/\mathbf{b})^{D-3}$ (see~\eqref{S14} for the numerical factors in the definition of $R_s$). The symbol $T(s,m_i,\mathbf{b})$ corresponds to all the non-divergent (in energy or mass) contributions to the amplitudes with any number of graviton exchanges. We have implicitly assumed that the eikonals behave as phases instead of operators since we are dealing with a purely elastic scenario in this paper. For a more general form of the equation above see equation (4.6) in \cite{Collado:2018isu}.

From these definitions and using observations from amplitude calculations we note that we can write the sum of leading contributions to the amplitudes with $n$ graviton exchanges as,
\begin{eqnarray}\label{eq:sumofleadingcont}
i \sum_{n=1}^{\infty} \tilde{\mathcal{A}}^{(1)}_{n}(s,m_i,\mathbf{b}) &=& i \tilde{\mathcal{A}}^{(1)}_{1}(s,m_i,\mathbf{b}) + i\tilde{\mathcal{A}}^{(1)}_{2}(s,m_i,\mathbf{b}) + \ldots \nonumber \\
&=& i \tilde{\mathcal{A}}^{(1)}_{1}(s,m_i,\mathbf{b}) + \frac{1}{2} \left(i\tilde{\mathcal{A}}^{(1)}_{1}(s,m_i,\mathbf{b})\right)^2 + \ldots \nonumber \\
&=& e^{i \delta^{(1)}(s,m_i,\mathbf{b})} - 1  \;,
\end{eqnarray}
where $\delta^{(1)}(s,m_i,\mathbf{b}) = \tilde{\mathcal{A}}^{(1)}_{1}(s,m_i,\mathbf{b})$ is the leading eikonal. Note that we refer to the leading contribution to each amplitude, with different numbers of graviton exchanges, via the superscript label $(1)$, each subleading order is then referred to by increasing the number in the superscript.

Expanding \eqref{eq:eiksmatrix} and collecting all the potential contributions at 2PM (i.e. at $\mathcal{O}(G_N^2)$) we can write an explicit expression for the first subleading eikonal,
\begin{eqnarray}
i \delta^{(2)}(s,m_i,\mathbf{b}) = i \tilde{\mathcal{A}}^{(2)}_{2} - i \tilde{\mathcal{A}}^{(1)}_{1} i \tilde{\mathcal{A}}^{(2)}_{1} = i \tilde{\mathcal{A}}^{(2)}_{2} \;, \label{eq:eiksubleading}
\end{eqnarray}
where in the second step we have used that in Einstein gravity we have $\mathcal{A}^{(2)}_{1}=0$ as can be seen from the results in section \ref{sec:treelevelamp}. Note that in other theories this may not be the case. For example, in supergravity we do find a contribution of the form $\mathcal{A}^{(2)}_{1}$ coming from the tree-level diagram with one RR field exchange between the scalar field and the stack of D-branes \cite{Collado:2018isu}.

Notice that for each eikonal we have,
\begin{equation}
i \delta^{(k)}(s,m_i,\mathbf{b}) \sim i \tilde{\mathcal{A}}^{(k)}_{k} \sim \mathcal{O}(G_N^{k})
\end{equation}
where $G_N$ is the usual Newton's constant. This relates the discussion presented here with the so called post-Minkowskian approximation discussed in \cite{Damour:2016gwp,Damour:2017zjx,Bjerrum-Bohr:2018xdl} and references therein. The leading eikonal corresponds to the 1PM order in the post-Minkowskian expansion, the subleading eikonal corresponds to the 2PM order and so on.

Using the various relations shown above and the results from section \ref{sec:amplitude} we can write the leading (1PM) and first subleading eikonals (2PM). Using equation \eqref{eq:1geamp} we find for the leading eikonal,
\begin{eqnarray}
\delta^{(1)}(s,m_i,\mathbf{b})\! =\!  \frac{16 \pi G_N \gamma (s)}{4Ep} \frac{\Gamma ( \frac{D-4}{2})}{4 \pi^{\frac{D-2}{2}} \mathbf{b}^{D-4}} \! =\!  \frac{ \pi G_N \Gamma ( \frac{D-4}{2})}{\pi^{\frac{D-2}{2}} \mathbf{b}^{D-4}}  \frac{(s-m_1^2 -m_2^2)^2 - \frac{4}{D-2} m_1^2 m_2^2}{\sqrt{(s-m_1^2 -m_2^2)^2 - 4 m_1^2 m_2^2 }}.
\label{S12}
\end{eqnarray}
We can verify the exponentiation of the eikonal at one-loop level by looking at the leading one-loop contribution, \eqref{eq:boxleading}, which we reproduce below,
\begin{eqnarray}
i \tilde{\mathcal{A}}^{(1)}_{2} = - \frac{\kappa_D^4 \gamma^2(s)}{(Ep)^2} \frac{1}{2^7 \pi^{D-2}} \Gamma^2 \left(\frac{D}{2}-2 \right) \frac{1}{\mathbf{b}^{2D-8}} = \frac{1}{2} (i \delta^{(1)})^2  \;.
\end{eqnarray}
Notice that this includes the appropriate numerical coefficient as required for the second line in \eqref{eq:sumofleadingcont} to hold.

Summing equations \eqref{eq:boxsubleadingb} and \eqref{eq:triips} we find for the subleading eikonal,
\begin{eqnarray}
&&\delta^{(2)}(s,m_i,\mathbf{b}) = 
\frac{(8 \pi G_N)^2 (m_1+m_2)}{Ep \, \pi^{D - \frac{3}{2}} }\, \frac{ \Gamma ( \frac{2D-7}{2}) \Gamma^2 ( \frac{D-3}{2})}{16 \,\mathbf{b}^{2D-7}} \nonumber \\
&& \times \left\{ \frac{ \gamma^2 (s)}{\Gamma (D-4) \left[ (s-m_1^2 -m_2^2)^2 - 4 m_1^2 m_2^2\right]}  + \frac{1}{4 \Gamma (D-3)}  \right. \nonumber \\
&&  \left. \times \left[ (s-m_1^2 -m_2^2)^2 - \frac{4m_1^2 m_2^2 }{(D-2)^2}  - \frac{(D-3) \left( (s-m_1^2 -m_2^2)^2 - 4 m_1^2 m_2^2 \right)  }{4 (D-2)^2} \right] \right\} \;. \nonumber \\
\label{S16}
\end{eqnarray}

The relation between the eikonal and the scattering angle relevant for discussing the post-Minkowskian expansion as well as comparing with results found using general relativity is given up to 2PM order by,
\begin{equation}
\theta = -\frac{1}{p}\frac{\partial}{\partial \mathbf{b}} \left( \delta^{(1)} + \delta^{(2)} \right) + \ldots  \label{eq:eikdefangle}
\end{equation}
where as previously stated $p$ is the absolute value of the space-like momentum in the center of mass frame of the two scattering particles.

\subsection{Various Probe Limits in Arbitrary $D$}\label{sec:probeind}

The corresponding deflection angle for the leading eikonal is given by,
\begin{eqnarray}
\theta^{(1)} = - \frac{1}{p} \frac{\partial}{\partial \mathbf{b}} \delta^{(1)} = \frac{4 \pi G_N \Gamma (\frac{D-2}{2}) \sqrt{s}}{\pi^{\frac{D-2}{2}} \mathbf{b}^{D-3}} \,\, \frac{(s-m_1^2 -m_2^2)^2 - \frac{4}{D-2} m_1^2 m_2^2}{(s-m_1^2 -m_2^2)^2 - 4 m_1^2 m_2^2} \;. 
\label{S13}
\end{eqnarray}
In the limit where both masses are zero (the ACV limit~\cite{Amati:1987wq}) the deflection angle can be written as follows,
\begin{eqnarray}
\theta^{(1)}_{ACV} = \frac{ \sqrt{\pi} \Gamma ( \frac{D}{2} )}{ \Gamma ( \frac{D-1}{2})} 
\left( \frac{R_s}{\mathbf{b}}\right)^{D-3}~;~~ 
R_s^{D-3} = \frac{16\pi G_N M^*}{(D-2)\Omega_{D-2}}~~;~~ 
\Omega_{D-2}=
\frac{2 \pi^{\frac{D-1}{2}}}{\Gamma ( \frac{D-1}{2})} \;,
\label{S14}
\end{eqnarray}
where $R_s$ is the effective Schwarzschild radius in $D$ dimensions, $M^*=\sqrt{s}$ or $M^*=m_1,m_2$ depending on which scale is larger, and $\Omega_{D-2}$ is the volume of a $(D-2)$-dimensional sphere. 

In the probe-limit with $m_2=0$ and $m_1=M$ where the mass, $M \gg \sqrt{s-M^2}$, we find that the deflection angle becomes,
\begin{eqnarray}
\theta^{(1)}_{\text{null}} = \frac{4 \pi G_N \Gamma (\frac{D-2}{2}) M}{\pi^{\frac{D-2}{2}} \mathbf{b}^{D-3}} \;,
\label{S15}
\end{eqnarray}
which is equal to the deflection angle that is obtained from the first term of the eikonal in equation (5.33) of \cite{Collado:2018isu} for $p=0$ and with the identification $N \tau_0=M$. This is also consistent with \eqref{eq:geogendnull} with the Schwarzschild radius defined as in \eqref{S14} with $M^*=M$.

In order to compare with the more general results for timelike geodesics in a $D$-dimensional Schwarzschild background obtained in section \ref{sec:geodesics} we can also take the timelike probe-limit. In this limit we have as before $m_1=M \gg m_2$ where $m_2 = m \neq 0$, so we have $\sqrt{s} \sim M$ and $(s-m_1^2 -m_2^2) \sim 2 E_2 M$. Using this we find,
\begin{equation}
\theta^{(1)}_{\text{timelike}} = \frac{\sqrt{\pi } \Gamma \left(\frac{D}{2}-1\right) \left((D-2) E_2^2-m^2\right)}{2 (E_2^2-m^2) \Gamma \left(\frac{D-1}{2}\right)} \left( \frac{R_s}{\mathbf{b}} \right)^{D-3} \;,
\label{s20}
\end{equation}
where we have used the definition of the Schwarzschild radius given in \eqref{S14}. This agrees with equations \eqref{eq:geoPhi} and \eqref{eq:geophi1} by using the relation, $J \simeq |p| |\mathbf{b}|$.

The subleading contribution to the deflection angle is given by using \eqref{eq:eikdefangle} and \eqref{S16},
\begin{eqnarray}
&& \theta^{(2)} = \frac{(8 \pi G_N)^2 (m_1+m_2)}{Ep^2 \pi^{D - \frac{3}{2}} }\, \frac{ 2\Gamma ( \frac{2D-5}{2}) \Gamma^2 ( \frac{D-3}{2})}{16 \,\mathbf{b}^{2D-6}} \nonumber \\
&& \times \left\{ \frac{ \gamma^2 (s)}{\Gamma (D-4) \left[ (s-m_1^2 -m_2^2)^2 - 4 m_1^2 m_2^2\right]}  + \frac{1}{4 \Gamma (D-3)}  \right. \nonumber \\
&&  \times \left. \left[ (s-m_1^2 -m_2^2)^2 - \frac{4m_1^2 m_2^2 }{(D-2)^2}  - \frac{(D-3) \left( (s-m_1^2 -m_2^2)^2 - 4 m_1^2 m_2^2 \right)  }{4 (D-2)^2} \right] \right\}  \;. \nonumber \\
\label{S17}
\end{eqnarray}
The subleading eikonal and deflection angle do not contribute in the limit when both masses are zero for any value of $D$ as discussed in \cite{Ciafaloni:2018uwe}. In the probe-limit where $m_2=0$ and $m_1 \equiv M \gg E_2$ we find for the subleading eikonal,
\begin{eqnarray}
&&\delta^{(2)}_{\text{null}} =  \frac{  (8\pi G_N M)^2   E_2 \, \Gamma ( \frac{2D-7}{2} ) 
\Gamma^2 ( \frac{D-3}{2})}{16 \pi^{D - \frac{3}{2}} \Gamma (D-4) \mathbf{b}^{2D-7} }  \nonumber \\
&& + \frac{  (8\pi G_N M)^2   E_2 \, \Gamma ( \frac{2D-7}{2} ) 
\Gamma^2 ( \frac{D-3}{2})}{16 \pi^{D - \frac{3}{2}} \Gamma (D-3)  \mathbf{b}^{2D-7} } 
\left( 1 - \frac{D-3}{4 (D-2)^2} \right) \;.
\label{S18}
\end{eqnarray}
The term in the first line is equal to the second term in the first line of equation (5.33) in \cite{Collado:2018isu} for $p=0$, while the term in the second line is equal to the sum of the terms in the third and fourth line of equation (5.33) for $p=0$. The corresponding deflection angle is given by,
\begin{eqnarray}
\theta^{(2)}_{\text{null}} &=&  \frac{  (8\pi G_N M)^2 \, \Gamma ( \frac{2D-5}{2} ) 
\Gamma^2 ( \frac{D-3}{2})}{8 \pi^{D - \frac{3}{2}} \Gamma (D-4) \mathbf{b}^{2D-6} }  \nonumber \\
&& + \frac{  (8\pi G_N M)^2 \, \Gamma ( \frac{2D-5}{2} ) 
\Gamma^2 ( \frac{D-3}{2})}{8 \pi^{D - \frac{3}{2}} \Gamma (D-3)  \mathbf{b}^{2D-6} } 
\left( 1 - \frac{D-3}{4 (D-2)^2} \right) \nonumber \\
&=& \frac{\sqrt{\pi } \Gamma \left(D-\frac{1}{2}\right)}{2 \Gamma (D-2)} \left( \frac{R_s}{\mathbf{b}} \right)^{2(D-3)} \;,
\label{S19}
\end{eqnarray}
where we have used the definition of the Schwarzschild radius given in \eqref{S14}. This agrees with equation \eqref{eq:geogendnull}. We can similarly look at the timelike probe-limit described before \eqref{s20}. In this case we find that \eqref{S17} becomes,
\begin{eqnarray}
\theta^{(2)}_{\text{timelike}} &=& \frac{\sqrt{\pi } \Gamma \left(D-\frac{5}{2}\right)}{\Gamma (D-2)} \frac{\left((2 D-5) (2 D-3) E_2^4+6 (5-2 D) E_2^2 m^2+3 m^4\right) }{8 (E_2^2-m^2)^2 } \left( \frac{R_s}{\mathbf{b}} \right)^{2(D-3)} \;. \nonumber \\
\end{eqnarray}
We can easily check, by using the relation $J \simeq |p| |\mathbf{b}|$, that this agrees with equations \eqref{eq:geoPhi} and \eqref{eq:geophi2} as expected.

\subsection{Various Probe Limits in $D=4$}\label{sec:probeind4}

We will now set $D=4$ in the various equations obtained in the previous subsection. We find that the leading eikonal in $D=4$ is equal to,
\begin{eqnarray}
\delta^{(1)} = - 2G_N \frac{2 \gamma (s)}{2Ep} \log \mathbf{b}  = - 2 G_N \frac{(s-m_1^2 -m_2^2)^2 - 2 m_1^2 m_2^2}{\sqrt{(s-m_1^2 -m_2^2)^2 - 4 m_1^2 m_2^2 }} \log \mathbf{b} \;,
\label{S1}
\end{eqnarray}
while the deflection angle is given by,
\begin{eqnarray}
\theta^{(1)} = - \frac{1}{p} \frac{\partial}{\partial \mathbf{b}} \delta^{(1)} = \frac{4 G_N \sqrt{s}}{\mathbf{b}}  \frac{(s-m_1^2 -m_2^2)^2 - 2 m_1^2 m_2^2}{(s-m_1^2 -m_2^2)^2 - 4 m_1^2 m_2^2} \;, 
\label{S2}
\end{eqnarray}
where we recall $E = \sqrt{s}$, $p$ is the absolute value of the three-dimensional momentum in the center of mass frame of the two scattering particles and we have used equations \eqref{eq:1gegms} and \eqref{eq:1geEP}. 

In the limit where both masses are zero (ACV limit) one gets,
\begin{eqnarray}
\delta_{ACV}^{(1)} = - 2 G_N s \,\,\log \mathbf{b} ~~;~~\theta_{ACV}^{(1)}
 = \frac{4G_N \sqrt{s}}{\mathbf{b}}= \frac{2 R_s}{\mathbf{b}}\;,
\label{S4}
\end{eqnarray}
which agrees with results found in \cite{Ciafaloni:2018uwe}. In the probe-limit where $m_2=0$ and $m_1=M$ we find the following eikonal,
\begin{eqnarray}
\delta_{\text{null}}^{(1)} = - 2 G_N (s-M^2) \log \mathbf{b}  \sim - 4 G_N M E_2 \log \mathbf{b} \;,
\label{S5}
\end{eqnarray}
where in the rest frame of the massive particle we have again used that $s-M^2 = 2ME_2$. Notice that equation \eqref{S5}  agrees with the first term of equation (5.41) in \cite{Collado:2018isu}. For the deflection angle we find,
\begin{eqnarray}
\theta_{\text{null}}^{(1)} = \frac{4 G_N \sqrt{s}}{\mathbf{b}} = \frac{4 G_N M}{\mathbf{b}} \;,
\label{S6}
\end{eqnarray}
when we assume that $M \gg E_2$. This agrees with the well known expression for the leading contribution to the deflection angle of a Schwarzschild black hole reproduced in \eqref{eq:geo4null}. Taking the timelike probe-limit of \eqref{S2} as described in the previous subsection where, $m_1=M \gg m_2$ where $m_2 = m \neq 0$, we find,
\begin{equation}
\theta_{\text{timelike}}^{(1)} = \frac{R_s (2E_2^2 - m^2)}{E_2^2-m^2}\frac{1}{\mathbf{b}} \;,
\end{equation}
which we find agrees with the first contribution to \eqref{eq:geogendtime4} as well as equivalent results in \cite{Damour:2017zjx}.

The subleading eikonal in $D=4$ is found to be,
\begin{eqnarray}
\delta^{(2)} &=& \frac{(8 \pi G_N)^2 (m_1+m_2)}{64 Ep \pi \mathbf{b}} \left[ \frac{15}{16} (s-m_1^2 -m_2^2)^2 - \frac{3}{4} m_1^2 m_2^2 \right] \nonumber \\
&=&  \frac{\pi G_N^2 (m_1+m_2)}{ 2\mathbf{b}  \sqrt{(s-m_1^2 -m_2^2)^2 - 4m_1^2 m_2^2} }
\left[ \frac{15}{4} (s-m_1^2 -m_2^2)^2 - 3 m_1^2 m_2^2 \right] \;.
\label{S7}
\end{eqnarray}
The factor of $m_1+m_2$ in front implies that the subleading eikonal in the massless  limit is vanishing~\cite{Amati:1990xe},
\begin{eqnarray}
\delta^{(2)}_{ACV} =0 ~~;~~~ \theta_{ACV}^{(2)} =0 \;.
\label{S8}
\end{eqnarray}
This also implies that there is no contribution of order $1/\mathbf{b}^2$ to the deflection angle which is consistent with the result found in the previous subsection that this contribution is zero for any number of spacetime dimensions. From equation \eqref{S7} we can compute the deflection angle,
\begin{eqnarray}
\theta^{(2)} =- \frac{1}{p} \frac{\partial}{\partial \mathbf{b}} \delta^{(2)} =  \frac{\pi G_N^2 (m_1+m_2) \sqrt{s}}{ \mathbf{b}^2 \left[ (s-m_1^2 -m_2^2)^2 - 4m_1^2 m_2^2\right] }
\left[ \frac{15}{4} (s-m_1^2 -m_2^2)^2 - 3 m_1^2 m_2^2 \right].
\label{S9}
\end{eqnarray}
In the probe-limit where $m_2 =0$ and $m_1=M$ we find,
\begin{eqnarray}
\delta_{\text{null}}^{(2)} = \frac{ 15 \pi G_N^2 M}{8\mathbf{b}} (s-M^2) \sim \frac{15 \pi (G_N M)^2E_2}{4\mathbf{b}} \;,
\label{S10}
\end{eqnarray}
which is equal to the second term of equation (5.41) in \cite{Collado:2018isu}. For the deflection angle we instead get,
\begin{eqnarray}
\theta_{\text{null}}^{(2)} =  \frac{15 \pi (G_N M)^2}{4\mathbf{b}^2}\;,
\label{S11}
\end{eqnarray}
which we find agrees with the subleading contribution found in \eqref{eq:geo4null}. We can also take the timelike probe-limit of \eqref{S9} for which we find,
\begin{equation}
\theta_{\text{timelike}}^{(2)} = \frac{3 \pi R_s^2 (5E^2_2 - m^2)}{16 (E_2^2- m^2)}\frac{1}{\mathbf{b}^2} \;.
\end{equation}
This agrees with the second contribution to \eqref{eq:geogendtime4} as well as equivalent results in \cite{Damour:2017zjx}.


\section{Discussion}
\label{sec:discussion}

In this paper we have studied the classical gravitational interaction between two massive scalars in $D$-dimensions up to 2PM order. As usual the spacetime dimension can be used as an infrared regulator and physical observables, such as the deflection angle discussed in section~\ref{sec:eikonal} have a smooth $D\to 4$ limit. The structure of the $D$-dimensional result is in some aspects richer than the one found in $D=4$. For instance the box integral provides not only the contribution necessary to exponentiate the leading energy behaviour of the tree-level diagram, but also a new genuine contribution to the subleading classical eikonal, see~\eqref{eq:boxsubleadingb}.

The box integral also provides a subsubleading contribution~\eqref{eq:boxsubsubleadingbq} that for $D\not=4$ has a new imaginary part, while its real part has a structure which also appears in the ${\cal O}(G_N^3)$ amplitude presented in~\cite{Bern:2019nnu}. In the ultra-relativistic limit $s\gg m_i^2$, this contribution is log-divergent and, if it does exponentiate as suggested in~\cite{Amati:1990xe}, it would provide a new {\em quantum} contribution, $\delta^{(2)}_{q}$, to the eikonal. For instance, from~\eqref{eq:boxsubsubleadingbq} in $D=4$ one would obtain\footnote{The triangle contributions discussed in section~\ref{sec:triangleamplitude} do not yield any log-divergent term.}
\begin{equation}
\delta^{(2)}_{q} \simeq  \frac{12 G_N^2 s}{\pi b^2} \log \frac{s}{m_1m_2} = 12 \frac{G_N s}{\hbar} \frac{\lambda_P^2}{\pi b^2} \log \frac{s}{m_1m_2}~,
\label{d2qua}
\end{equation}
where we have taken the limit $s\gg m_i^2$ in order to compare with equation~(5.18) of\footnote{That equation should have an extra factor of $\lambda_P$ and $\delta^{(n)}_{\rm here}= 2\delta^{(n-1)}_{\rm there}$.}~\cite{Amati:1990xe} and $\lambda_P$ is the Planck length. By restoring the factors of $\hbar$ we can see from~\eqref{S1} that $\delta^{(1)}/\hbar$ is dimensionless and is therefore the combination that is exponentiated. On the contrary, $\delta_q^{(2)}$ is dimensionless without the need for any factor of $\hbar$ (see the first expression in~(4.1) or equivalently it can be written in terms of $\lambda_P^2$ if we extract a factor of $1/\hbar$), which highlights its quantum nature.

An interesting feature of the 2PM eikonal phase for massive scalars that we have obtained is that its ultra-relativistic limit smoothly reproduces the massless result up to 2PM order. This is valid also for the quantum contribution mentioned above~\eqref{d2qua}. By comparing the results of~\cite{Amati:1990xe} and~\cite{Bern:2019nnu}, the same property does not seem to hold in the 3PM case and it would be very interesting to understand the origin of this mismatch. Another interesting development would be to generalise the analytic bootstrap approach of~\cite{Amati:1990xe,Ciafaloni:2018uwe} beyond the massless $D=4$ case. In that approach the quantum part of the eikonal $\delta^{(2)}_q$ plays an important role in the derivation of the subsequent {\em classical} PM order and we expect that a similar pattern is valid also beyond the setup of~\cite{Amati:1990xe,Ciafaloni:2018uwe}. This approach has the potential to provide an independent derivation of the 3PM eikonal phase both in the massless higher dimensional case and in the physically interesting case of the massive scattering in $D=4$. 

\section*{Acknowledgements}

We would like to thank Z. Bern, N.E.J. Bjerrum-Bohr, A. Brandhuber, P.H. Damgaard,  T. Damour, G. Travaglini, P. Vanhove, J. Vines, C. Wen, C. White and Z. Zahraee for discussions, M. Ciafaloni, D. Colferai and G. Veneziano for extended  correspondence and many clarifications, M. Bianchi for a critical reading of the manuscript and S. Thomas for discussions and collaboration on related projects. This work was supported in part by the Science and Technology Facilities Council (STFC) Consolidated Grant ST/L000415/1 {\it String theory, gauge theory \& duality}. AKC is supported by an STFC studentship.


\appendix

\section{High-Energy Expansion of Various Integrals} \label{app:integrals}

In the classical regime the centre of mass energy $\sqrt{s}$ and the masses $m_i^2$ are much larger than the momentum exchanged $q$. In this limit the integrals appearing in the amplitudes discussed in the main text can be performed so as to extract the leading and the subleading contributions (we also calculate the subsubleading contribution to the scalar box integral). Our approach is the following: we first write our starting point in terms of Schwinger parameters $t_i$, then we perform the integrals over the $t$'s parametrising the scalar propagators by using a saddle point approximation, finally the integrals over the graviton propagators reduce to those of an effective two-point function. In the following subsection we give a detailed analysis of the so called scalar box integral, showing that for a general spacetime dimension $D$ it provides a classical contribution proportional to $(D-4)$. We will also give results for the triangle integrals which are necessary to evaluate the full classical contribution at one-loop.

\subsection{Scalar Box Integral} \label{app:boxintegrals}

In this subsection we will discuss the scalar box integral. To start we will be evaluating,
\begin{equation}
  \label{eq:1lbmi}
  \mathcal{I}_4(s,t) = \int \!\frac{d^Dk}{(2\pi)^D}\, \frac{1}{k^2} \,\frac{1}{(k+k_1)^2+m_1^2} \,\frac{1}{(q+k)^2} \, \frac{1}{(k_2-k)^2+m_2^2} \;.
\end{equation}
After a Wick rotation and introducing Schwinger parameters the integral over the loop momentum $k$ is Gaussian and can be readily performed. After evaluating this we find,
\begin{equation}
\mathcal{I}_4(s,t) = i \int_{0}^{\infty} \prod_{i=1}^{4}  \, d t_i \frac{T^{-\frac{D}{2}}}{(4\pi)^{\frac{D}{2}}} \exp \left[-\frac{2 k_1 k_2 t_2 t_4 + q^2 t_1 t_3 + t_2^2 m_1^2 + t_4^2 m_2^2}{T} \right]\;,
\end{equation}
where $q\equiv k_1+k_3$ is the momentum exchanged and we have defined $T=\sum_{i} t_i$. The form of the equation above is suggestive because we are interested in the limit where $|k_1 k_2|, m_i^2 \gg q^2$, this means that the integral over $t_2,t_4$ can be performed with a saddle point approximation around $t_2=t_4=0$.

What makes this integral awkward is that its region of integration is just the positive quadrant in $t_2,t_4$. In order to circumvent this problem it is convenient to sum the contribution of the crossed box integral. In terms of Schwinger parameters this is given by
\begin{equation}
\mathcal{I}_4(u,t) = i \int_{0}^{\infty} \prod_{i=1}^{4}  \, d t_i \frac{T^{-\frac{D}{2}}}{(4\pi)^{\frac{D}{2}}} \exp \left[-\frac{2 k_2 k_3 t_2 t_4 + q^2 t_1 t_3 + t_2^2 m_1^2 + t_4^2 m_2^2}{T} \right] \;.
\end{equation}
Notice that $\mathcal{I}_4(u,t)$ can be obtained from $\mathcal{I}_4(s,t)$ by swapping $k_1 \leftrightarrow k_3$. In order to combine $\mathcal{I}_4(s,t)$ and $\mathcal{I}_4(u,t)$, it is convenient to define,
\begin{equation}
  \tilde{k}^2 = k_1 k_2 + \frac{q^2}{4} = - k_2 k_3 - \frac{q^2}{4} \;.
\end{equation}
Then we can rewrite,
\begin{equation}
  \label{eq:intco}
  \mathcal{I}_4(s,t) = i \int_{0}^{\infty} \prod_{i=1}^{4}  \, d t_i f(\tilde{k}^2,t_i)\;,~~~ \mathcal{I}_4(u,t) = i \int_{0}^{\infty} \prod_{i=1}^{4}  \, d t_i f(-\tilde{k}^2,t_i)\;,~~~
\end{equation}
where,
\begin{equation}\label{eq:fuf}
  f(\tilde{k}^2,t_i) =  \frac{e^{-q^2 \frac{t_1 t_3}{T}}}{(4\pi)^{\frac{D}{2}}\,T^{\frac{D}{2}}} \exp \left[  -\frac{(t_2\; t_4)}{T} \left(\begin{matrix} m_1^2 & \tilde{k}^2 \\ \tilde{k}^2 & m_2^2 \end{matrix} \right) \left(\begin{matrix} t_2 \\ t_4 \end{matrix} \right)  \right]
   e^{\frac{q^2}{2T} |t_2 t_4|}\;. 
\end{equation}
As previously mentioned we are interested in performing these integrals in the limit where $|k_1 k_2|, m_i^2,1/t_2,1/t_4$ are all of the same order and much bigger than $q^2$. We can therefore Taylor expand the integrands for small $t_2$ and $t_4$ and at the leading order we simply obtain the function~\eqref{eq:fuf} where $T$ reduces to $t_1+t_3$ and the last exponential can be neglected. It is therefore convenient to define,
\begin{equation}
T_0 = t_1 + t_3\;.
\end{equation}
By expressing the two integrals in this way, we can see that they are equivalent under the change $t_2 \rightarrow - t_2$ or $t_4 \rightarrow - t_4$. We note that, $I_4(t_2, -t_4) = J_4(t_2, t_4)$, $I_4(-t_2, t_4) = J_4(t_2, t_4)$, $I_4(-t_2, -t_4) = I_4(t_2, t_4)$ where $I_4, J_4$ are the integrands of $\mathcal{I}_4(s,t), \mathcal{I}_4(u,t)$ respectively. We can therefore write the combination of box and crossed box integrals as,
\begin{eqnarray} \label{eq:i4j4full}
\mathcal{I}_4(s,t) + \mathcal{I}_4(u,t) &=& \frac{i}{2} \int_{0}^{\infty} \, d t_1 d t_3 \int_{-\infty}^{\infty} \, d t_2 d t_4 \frac{e^{-q^2 \frac{t_1 t_3}{T}}}{(4\pi)^{\frac{D}{2}}\,T^{\frac{D}{2}}} \nonumber \\
&& \qquad \times \exp \left[  -\frac{(t_2\; t_4)}{T} \left(\begin{matrix} m_1^2 & \tilde{k}^2 \\ \tilde{k}^2 & m_2^2 \end{matrix} \right) \left(\begin{matrix} t_2 \\ t_4 \end{matrix} \right)  \right] e^{\frac{q^2}{2T} |t_2 t_4|}\;,
\end{eqnarray}
where $T$ is now explicitly defined as $T=T_0+|t_2|+|t_4|$. Note that we have written some quantities as $|t_2|, |t_4|$ since the original domain of integration is for $t_{2,4} \geq 0$.

Expanding \eqref{eq:i4j4full} around $(t_2,t_4)=(0,0)$ the leading contribution (i.e. the one that eikonalises the tree-level amplitude, see comments below \eqref{S12}), which we denote as $\mathcal{I}_4^{(1)}(s,t) + \mathcal{I}_4^{(1)}(u,t)$, can be written as a Gaussian integral,
\begin{eqnarray}\label{eq:i4j4leadingint}
 \mathcal{I}_4^{(1)}(s,t) + \mathcal{I}_4^{(1)}(u,t) &=& i \int_{0}^{\infty}\!\! dT_0\, \frac{T_0^{1-\frac{D}{2}}}{(4\pi)^{\frac{D}{2}}} \int_{0}^{1} d x_1 \, \exp \left[- q^2 x_1 (1-x_1) T_0 \right] \nonumber \\
&&\times  \frac{1}{2} \int_{-\infty}^\infty dt_2 \, dt_4 \, \exp \left[  -\frac{(t_2\; t_4)}{T_0} \left(\begin{matrix} m_1^2 & \tilde{k}^2 \\ \tilde{k}^2 & m_2^2 \end{matrix} \right) \left(\begin{matrix} t_2 \\ t_4 \end{matrix} \right) \right]\;,  
\end{eqnarray}
where $x_1=t_1/T_0$. The quadrants with $t_2, t_4>0$ and $t_2,t_4<0$ yield the same contribution corresponding $\mathcal{I}_4^{(1)}(s,t)$, while those with $t_2>0>t_4$ and $t_4>0>t_2$ are again identical and correspond to $\mathcal{I}_4^{(1)}(u,t)$. 

We should also recall that $\tilde{k}^2$ is not a kinematic variable directly relevant to the amplitude calculations. In the resulting expression from performing the Gaussian integral over $t_2, t_4$ in \eqref{eq:i4j4leadingint} above we need to be careful and also substitute for,
\begin{equation}\label{eq:tildekexp}
\tilde{k}^2 = k_1 k_2 + \frac{q^2}{4} = k_1 k_2 \left(1 + \frac{q^2}{4 k_1 k_2} \right) \;, 
\end{equation}
whilst taking into account that we are interested in the limit where $|k_1 k_2|, m_i^2 \gg q^2$. The remaining two integrals over $T_0$ and $x_1$ can be decoupled and, by collecting everything together, we find
\begin{equation}
\mathcal{I}_4^{(1)}(s,t)  + \mathcal{I}_4^{(1)}(u,t) = \frac{1}{2} \frac{\pi}{(4\pi)^{\frac{D}{2}}} \frac{-1}{\sqrt{(k_1 k_2)^2 - m_1^2 m_2^2}}\,\Gamma \left(\frac{6-D}{2}\right)\, \frac{\Gamma^2\!\left(\frac{D-4}{2}\right)}{\Gamma(D-4)} (q^2)^{\frac{D-6}{2}} \;.
\label{eq:i4leading}
\end{equation}
Since we will look at the resulting amplitudes in impact parameter space it is worth calculating the expressions above in impact parameter space. Using \eqref{eq:ampips} and \eqref{eq:impoformu} we find that \eqref{eq:i4leading} becomes\footnote{Note that in this appendix we are not including the normalisation factor, $1/4Ep$, found in \eqref{eq:ampips}.},
\begin{equation}
\tilde{\mathcal{I}}_{4}^{(1)}(s,t) + \tilde{\mathcal{I}}_{4}^{(1)}(u,t) = \frac{-1}{2^7 \pi^{D-2}} \frac{1}{\sqrt{(k_1 k_2)^2 - m_1^2 m_2^2}} \Gamma^2\!\left(\frac{D}{2}-2\right) \frac{1}{\mathbf{b}^{2D-8}} \;.
\end{equation}
We can see that in $D=4$ the result is IR divergent and dimensional regularisation can be used to extract the $\log \mathbf{b}^2$ term we are interested in. In order to implement this we  would use,
\begin{eqnarray}
\frac{\Gamma ( \frac{D-4}{2})}{4 \pi^{\frac{D-2}{2}}\mathbf{b}^{D-4}} \Longrightarrow - \frac{1}{2\pi} \log \mathbf{b} \;.
\end{eqnarray}
Note that if we were to directly integrate over one of the quadrants in order to calculate the result for just one of the scalar box diagrams we find at leading order,
\begin{eqnarray}\label{eq:1lb1q}
\mathcal{I}_4^{(1)}(u,t) &=& \frac{i}{(4\pi)^{\frac{D}{2}}} \frac{\ln\left[\left(\frac{\sqrt{k_2 k_3 + m_1 m_2}+\sqrt{k_2 k_3 - m_1 m_2}}{\sqrt{2 m_1 m_2}}\right)^2\right]}{2 \sqrt{(k_2k_3)^2 - m_1^2 m_2^2}}\,\Gamma \left(\frac{6-D}{2}\right)\, \frac{\Gamma^2\!\left(\frac{D-4}{2}\right)}{\Gamma(D-4)} (q^2)^{\frac{D-6}{2}} \nonumber \\
& \approx & \frac{i}{(4\pi)^{\frac{D}{2}}} \frac{\text{arcsinh}\left(\sqrt{\frac{\sigma-1}{2}}\right)}{ \sqrt{(k_1k_2)^2 - m_1^2 m_2^2}}\,\Gamma \left(\frac{6-D}{2}\right)\, \frac{\Gamma^2\!\left(\frac{D-4}{2}\right)}{\Gamma(D-4)} (q^2)^{\frac{D-6}{2}} \;,
\end{eqnarray}
where we have used $-k_2 k_3 = k_1 k_2 + q^2/2$ in the second line to express this in terms of $k_1 k_2$ and as we defined in the main text $\sigma = \frac{- k_1 k_2}{m_1 m_2}$. Note that to find the equivalent expression for $\mathcal{I}_4^{(1)}(s,t)$ we switch $k_3 \leftrightarrow k_1$ in the first line above.

Expanding \eqref{eq:i4j4full} further we find the following subleading contribution,
\begin{eqnarray} \label{eq:ij4subleading}
\mathcal{I}_4^{(2)}(s,t) + \mathcal{I}_4^{(2)}(u,t) &=& i \int_{0}^{\infty}\!\! dT_0\, \frac{T_0^{1-\frac{D}{2}}}{(4\pi)^{\frac{D}{2}}} \int_{0}^{1} d x_1 \, \exp \left[- q^2 x_1 (1-x_1) T_0 \right] \nonumber \\
&&\times  \frac{1}{2} \int_{-\infty}^\infty dt_2 \, dt_4 \, \exp \left[  -\frac{(t_2\; t_4)}{T_0} \left(\begin{matrix} m_1^2 & \tilde{k}^2 \\ \tilde{k}^2 & m_2^2 \end{matrix} \right) \left(\begin{matrix} t_2 \\ t_4 \end{matrix} \right) \right] \nonumber \\
&&\times \frac{|t_2|+|t_4|}{T_0^2} \left(2 k_1 k_2 t_2 t_4+m_1^2 t_2^2+m_2^2 t_4^2+q^2 T_0^2 x_1 (1-x_1) - \frac{D}{2} T_0 \right) \nonumber \\
&=& \frac{i\sqrt{\pi}}{2 (4\pi)^{\frac{D}{2}}} \frac{m_1+m_2}{(k_1 k_2)^2 - m_1^2 m_2^2} \,\Gamma \left(\frac{5-D}{2}\right)\,\frac{\Gamma^2\left(\frac{D-3}{2}\right)}{\Gamma(D-4)} (q^2)^{\frac{D-5}{2}}\;, 
\end{eqnarray}
where we have also taken into account the fact that we need to substitute for $\tilde{k}^2$ using \eqref{eq:tildekexp} and find the leading contribution in $|k_1 k_2|, m_i^2 \gg q^2$ after performing the substitution. It is worth looking at the impact parameter space expression of the above result. In impact parameter space we have,
\begin{equation}
\tilde{\mathcal{I}}_4^{(2)}(s,t) + \tilde{\mathcal{I}}_4^{(2)}(u,t) = \frac{i}{2^5 \pi^{D-\frac{3}{2}}} \frac{m_1+m_2}{(k_1 k_2)^2 - m_1^2 m_2^2} \,\Gamma \left(\frac{2D-7}{2}\right)\,\frac{\Gamma^2\left(\frac{D-3}{2}\right)}{\Gamma(D-4)} \frac{1}{(\mathbf{b}^2)^{D-\frac{7}{2}}} \;,
\end{equation}
which we see vanishes for $D=4$.

The subsubleading integral is slightly more nuanced. In this case not only do we need to resolve the third term in the expansion of \eqref{eq:i4j4full} but we also need to take into account the contribution coming from the expansion of $\tilde{k}^2$ in the result for \eqref{eq:i4j4leadingint}. The extra contribution from \eqref{eq:i4j4leadingint} is given by,
\begin{equation} \label{eq:xtraconti4ltoi4ssl}
\frac{i}{2} \frac{\pi}{(4\pi)^{\frac{D}{2}}} \frac{k_1 k_2}{4[m_1^2 m_2^2-(k_1 k_2)^2]^{3/2}}\,\Gamma \left(\frac{6-D}{2}\right)\, \frac{\Gamma^2\!\left(\frac{D-4}{2}\right)}{\Gamma(D-4)} (q^2)^{\frac{D-4}{2}} \;.
\end{equation}
From the expansion of \eqref{eq:i4j4full} to subsubleading order we find,
\begin{eqnarray}
\mathcal{I}_4^{(3)}(s,t) + \mathcal{I}_4^{(3)}(u,t) &=& i \int_{0}^{\infty}\!\! dT_0\, \frac{T_0^{1-\frac{D}{2}}}{(4\pi)^{\frac{D}{2}}} \int_{0}^{1} d x_1 \, \exp \left[- q^2 x_1 (1-x_1) T_0 \right] \nonumber \\
&&\times  \frac{1}{2} \int_{-\infty}^\infty dt_2 \, dt_4 \, \exp \left[  -\frac{(t_2\; t_4)}{T_0} \left(\begin{matrix} m_1^2 & \tilde{k}^2 \\ \tilde{k}^2 & m_2^2 \end{matrix} \right) \left(\begin{matrix} t_2 \\ t_4 \end{matrix} \right) \right] \nonumber \\
&&\times \frac{1}{8 T_0^4} \biggl\{ (|t_2|+|t_4|)^2 \biggl[-4 D T_0 \left(t_4 \left(2 \tilde{k}^2 t_2+m_2^2 t_4\right)+m_1^2 t_2^2+q^2 t_1 t_3\right) \nonumber \\
&& +(D+2) D T_0^2+4 \left(t_4 \left(2 \tilde{k}^2 t_2+m_2^2 t_4\right)+m_1^2 t_2^2+q^2 t_1 t_3\right) \\ \nonumber
&& \times \left(2 \tilde{k}^2 t_2 t_4+m_1^2 t_2^2+m_2^2 t_4^2+q^2 t_1 t_3-2 T_0 \right) \biggr] + 4 T_0^3 q^2 |t_2| |t_4| \biggr\} \;,
\end{eqnarray}
where to solve the various resulting integrals in the Gaussian integration over $t_2,t_4$ we refer to the building blocks computed in appendix \ref{app:aux}. Taking into account the contribution \eqref{eq:xtraconti4ltoi4ssl} and again substituting as per \eqref{eq:tildekexp} we have for the final result,
\begin{eqnarray}
&& \mathcal{I}_4^{(3)}(s,t) + \mathcal{I}_4^{(3)}(u,t) = \frac{i}{8(4\pi)^{\frac{D}{2}}} \Gamma \left(\frac{4-D}{2}\right)\, \frac{\Gamma^2\!\left(\frac{D-2}{2}\right)}{\Gamma(D-4)}  (q^2)^{\frac{D-4}{2}} \frac{1}{D-4} \nonumber \\
&& \times \biggl[ \frac{4(5-D)}{(k_1 k_2)^2 - m_1^2 m_2^2} \left( 1 + \frac{2k_1 k_2}{[(k_1k_2)^2 - m_1^2 m_2^2]^{1/2}} \text{arcsinh} \left( \sqrt{\frac{\sigma-1}{2}} \right) \right) \nonumber \\
&& + i \frac{\pi (D-4) (k_1+k_2)^2}{[(k_1k_2)^2 - m_1^2 m_2^2]^{3/2}} \biggr]  \;,
\end{eqnarray}
where as we defined in the main text, $\sigma = \frac{- k_1 k_2}{m_1 m_2}$. A curious feature of the expression above is that it is purely imaginary for $D=4$ since the last line representing the real component vanishes. This expression has been checked against known result in $D=4$ found in \cite{Ellis:2007qk}. In impact parameter space the above expression reads,
\begin{eqnarray}
&& \tilde{\mathcal{I}}_4^{(3)}(s,t) + \tilde{\mathcal{I}}_4^{(3)}(u,t) = \frac{i}{2^7 \pi^{D-1}} \Gamma^2\!\left(\frac{D-2}{2}\right) \frac{1}{(\mathbf{b}^2)^{D-3}} \nonumber \\
&& \times \biggl[ \frac{4(5-D)}{(k_1 k_2)^2 - m_1^2 m_2^2} \left( 1 + \frac{2k_1 k_2}{[(k_1k_2)^2 - m_1^2 m_2^2]^{1/2}} \text{arcsinh} \left( \sqrt{\frac{\sigma-1}{2}} \right) \right) \nonumber \\
&& + i \frac{\pi (D-4) (k_1+k_2)^2}{[(k_1k_2)^2 - m_1^2 m_2^2]^{3/2}} \biggr] \;.
\end{eqnarray}

\subsection{Triangle Integrals} \label{app:triintegrals}
In this subsection we will derive results for the integrals relevant for the triangle amplitudes. The first integral we need to calculate is,
\begin{eqnarray}
\mathcal{I}_{3}(m_i) &=& \int \!\frac{d^Dk}{(2\pi)^D}\, \frac{1}{k^2} \,\frac{1}{(q+k)^2} \,\frac{1}{(k+k_i)^2+m_i^2}  \;.
\end{eqnarray}
As in the previous subsection we can write this integral in terms of Schwinger parameters and perform the Gaussian integral over the loop momenta. This yields,
\begin{equation}
\mathcal{I}_3 = i \int_{0}^{\infty} \prod_{i=1}^{3}  \, d t_i \frac{T^{-\frac{D}{2}}}{(4\pi)^{\frac{D}{2}}} \exp \left[-\frac{ m_i^2 t_3^2 + q^2 t_1 t_2}{T} \right]\;, \label{eq:i3full}
\end{equation}
where $q\equiv k_1+k_3$ is the momentum exchanged and $T=\sum_{i} t_i$. We have written this integral in a suggestive way because we are interested in the limit where $m_i^2 \gg q^2$, this means that the integral over $t_3$ can be performed with a saddle point approximation around $t_3=0$. To make it easier to perform the relevant expansion we write $T=T_0+| t_3 |$ where $T_0=t_1+t_2$. Doing so we find at leading order,
\begin{eqnarray}
\mathcal{I}_{3}^{(1)}(m_i) &=& i \int_{0}^{\infty} d T_0 \frac{T_0^{-\frac{D}{2}}}{(4\pi)^{\frac{D}{2}}} \int_{0}^{1} dx_2 \exp \left[- q^2 T_0 x_2(1-x_2) \right] \int_{0}^{\infty} \exp \left[- \frac{m_i^2 t_3^2}{T_0} \right] \nonumber \\
&=& \frac{i}{(4 \pi)^{\frac{D}{2}}} \frac{\sqrt{\pi}}{2 m_i} \Gamma \left(\frac{5-D}{2} \right)\, \frac{\Gamma^2\!\left(\frac{D-3}{2}\right)}{\Gamma(D-3)} (q^2)^{\frac{D-5}{2}} \;, \label{eq:i3m1}
\end{eqnarray}
where we have written $x_2=t_2/T_0$. Expanding \eqref{eq:i3full} further we find for the subleading contribution,
\begin{eqnarray}
\mathcal{I}_{3}^{(2)}(m_i) &=& i \int_{0}^{\infty} d T_0 \frac{T_0^{-\frac{D}{2}}}{(4\pi)^{\frac{D}{2}}} \int_{0}^{1} dx_2 \exp \left[- q^2 T_0 x_2(1-x_2) \right] \int_{0}^{\infty} \exp \left[- \frac{m_i^2 t_3^2}{T_0} \right] \nonumber \\
&& \quad \times \frac{1}{2 T_0^2} |t_3| \left( 2 q^2 t_1 t_2 - D T_0 + 2 m_i^2 t_3^2 \right) \nonumber \\
&=& -\frac{i}{(4 \pi)^{\frac{D}{2}}} \frac{1}{2 m_i^2} \Gamma \left(\frac{4-D}{2} \right) \, \frac{\Gamma^2\!\left(\frac{D-2}{2}\right)}{\Gamma(D-3)} (q^2)^{\frac{D-4}{2}} \;. \label{eq:i3m1sub}
\end{eqnarray}
The equations above in impact parameter space read,
\begin{eqnarray}
\tilde{\mathcal{I}}_{3}^{(1)}(m_i) &=& \frac{i}{(\pi)^{D-\frac{3}{2}}} \frac{\sqrt{\pi}}{64 m_i} \Gamma \left(\frac{2D-7}{2} \right)\, \frac{\Gamma^2\!\left(\frac{D-3}{2}\right)}{\Gamma(D-3)} \frac{1}{(\mathbf{b}^2)^{D-\frac{7}{2}}} \;, \label{eq:i3m1ips}
\end{eqnarray}
and,
\begin{eqnarray}
\tilde{\mathcal{I}}_{3}^{(2)}(m_i) &=& -\frac{i}{\pi^{D-1}} \frac{1}{32 m_i^2} \, \Gamma^2\!\left(\frac{D-2}{2}\right) \frac{1}{(\mathbf{b}^2)^{D-3}} \;. \label{eq:i3m1subips}
\end{eqnarray}

We also want to consider the integral given by,
\begin{equation}
\mathcal{I}_{3}^{\mu \nu}(m_i) = \int \!\frac{d^Dk}{(2\pi)^D}\, \frac{1}{k^2}  \,\frac{1}{(q+k)^2} \, \frac{1}{(k_i+k)^2 + m_i^2} k^{\mu} k^{\nu} \;.
\end{equation}
Employing Schwinger parameters as before we find that,
\begin{equation}
\mathcal{I}_3 = i \int_{0}^{\infty} \prod_{i=1}^{3}  \, d t_i \frac{T^{-\frac{D}{2}}}{(4\pi)^{\frac{D}{2}}} \exp \left[-\frac{ m_i^2 t_3^2 + q^2 t_1 t_2}{T} \right] \left( \frac{1}{2T} \eta^{\mu \nu} + \frac{1}{T^2}(q t_2 + k_i t_3)^{\mu}(q t_2 + k_i t_3)^{\nu} \right) \;, \label{eq:imunu3full}
\end{equation}
where the various symbols have been previously defined. Using the same method as for the previous two integrals we find the following result at leading order,
\begin{eqnarray}
\mathcal{I}_{3}^{(1)\,\mu \nu}(m_i) &=& \frac{i}{4m_i} \frac{1}{(4\pi)^{\frac{D}{2}}}\biggl[ (q^2)^{\frac{D-3}{2}} \sqrt{\pi} \frac{\Gamma{\left( \frac{3-D}{2} \right)} \Gamma^2{\left( \frac{D-1}{2} \right)}}{\Gamma{\left( D-1 \right)}} \left( \eta^{\mu \nu} + \frac{k_i^{\mu} k_i^{\nu}}{m_i^2} \right. \nonumber \\
&& \left. - (D-1)\frac{q^{\mu}q^{\nu}}{q^2} \right) + 2(q^2)^{\frac{D-4}{2}} \frac{\Gamma{\left( \frac{4-D}{2} \right)} \Gamma{\left( \frac{D-2}{2} \right)}\Gamma{\left( \frac{D}{2} \right)}}{\Gamma{\left( D-1 \right)}} \frac{q^{(\mu}k_i^{\nu)}}{m_i} \biggr] \;. \label{eq:i3munuleading}
\end{eqnarray}
Although this is the result at leading order in the expansion around the saddle point $t_3=0$ we can identify the second line as subleading contributions to the integral in the limit given by \eqref{eq:heml}. We can see this by looking at how a contraction between $q$ and an external momenta behaves,
\begin{equation}
k_1^{\mu} q_{\mu} = k_1^{\mu} (k_{1 \mu}+k_{3 \mu}) = \frac{1}{2} (k_1+k_3)^2 = \frac{1}{2}q^2
\end{equation}
where we have used the fact that $k_1^2=k_3^2$. Power counting with the above relation identifies the last line of \eqref{eq:i3munuleading} as subleading. This type of argument extends to contractions between any external momenta and $q$ since we can always write $q=k_1+k_3=-k_2-k_4$.

For the next order in the expansion around the saddle point we have,
\begin{eqnarray}
\mathcal{I}_{3}^{(2)\,\mu \nu}(m_i) &=& \frac{-i}{(4\pi)^{\frac{D}{2}}} \frac{1}{4m_i^2} \biggl[ (q^2)^{\frac{D-2}{2}} \frac{\Gamma{\left( \frac{2-D}{2} \right)} \Gamma^2{\left( \frac{D}{2} \right)}}{\Gamma{\left( D-1 \right)}} \left( \eta^{\mu \nu} + \frac{2 k_i^{\mu} k_i^{\nu}}{m_i^2} \right. \nonumber \\
&& \left. - D \frac{q^{\mu}q^{\nu}}{q^2} \right) + \frac{\sqrt{\pi}(D-1)}{2}(q^2)^{\frac{D-3}{2}} \frac{\Gamma{\left( \frac{3-D}{2} \right)} \Gamma^2{\left( \frac{D-1}{2} \right)}}{\Gamma{\left( D-1 \right)}} \frac{q^{(\mu} k_i^{\nu)}}{m_i} \biggr] \;. \label{eq:i3munusubleading}
\end{eqnarray}
We note that as in \eqref{eq:i3munuleading} the second line above is kinematically subleading with respect to the first line.

We can write the above expressions in impact parameter space as we have done with previous results. To make these expressions clear we will write them after we have contracted with external momenta. So we have,
\begin{eqnarray}\label{eq:i3munuleadingips}
k_{j \, \mu} k_{j \, \nu} \tilde{\mathcal{I}}_{3}^{(1)\,\mu \nu}(m_i) &=& \frac{i}{32m_i} \frac{1}{\pi^{D-\frac{3}{2}}} \biggl[ \frac{1}{(\mathbf{b}^2)^{D-\frac{5}{2}}} \frac{\Gamma{\left( \frac{2D-5}{2} \right)} \Gamma^2{\left( \frac{D-1}{2} \right)}}{\Gamma{\left( D-1 \right)}} \left( -m_j^2 + \frac{(k_i k_j)^2}{m_i^2} \right. \nonumber \\
&& \left. - (D-1) \frac{-2D^2 +7D +5}{4\mathbf{b}^2} \right) + (-1)^{j+1} \frac{k_i k_j}{m_i} \frac{2}{(\mathbf{b}^2)^{D-2}} \frac{\csc \left( \frac{\pi D}{2} \right)  \Gamma{\left( \frac{D}{2} \right)}}{\Gamma{\left( \frac{4-D}{2} \right)}} \biggr] \;, \nonumber \\
\end{eqnarray}
and for the subleading expression,
\begin{eqnarray}
k_{j \, \mu} k_{j \, \nu} \tilde{\mathcal{I}}_{3}^{(2)\,\mu \nu}(m_i) &=& \frac{-i}{16 m_i^2} \frac{1}{\pi^{D-1}} \biggl[ \frac{1}{(\mathbf{b}^2)^{D-2}} \frac{\Gamma^2{\left( \frac{D}{2} \right)}}{D-2} \left( -m_j^2 + \frac{2(k_i k_j)^2}{m_i^2} + \frac{D^2(D-2)}{2\mathbf{b}^2} \right) \nonumber \\
&& - (-1)^{j+1} \frac{k_i k_j}{m_i} \frac{2 \pi^{\frac{3}{2}}}{(\mathbf{b}^2)^{D-\frac{3}{2}}} \frac{\Gamma{\left( \frac{2D-3}{2} \right)} \Gamma{\left( \frac{D+1}{2} \right)} \sec \left( \frac{\pi D}{2} \right)}{\Gamma{\left( D-1 \right)} \Gamma{\left( \frac{1-D}{2} \right)} } \biggr] \;. \label{eq:i3munusubleadingips}
\end{eqnarray}

\subsection{Auxiliary Integrals} \label{app:aux}
In this appendix we will give results for various sub-integrals which appear in appendix \ref{app:integrals}. We find integrals such as,
\begin{eqnarray}
\mathcal{\hat{I}}^{(a)}_{t_2} &=& \int_{-\infty}^{\infty} dt_2 \, dt_4 |t_2|^{2m} t_2^{2n} \exp \left[  -\frac{(t_2\; t_4)}{T_0} \left(\begin{matrix} m_1^2 & \tilde{k}^2 \\ \tilde{k}^2 & m_2^2 \end{matrix} \right) \left(\begin{matrix} t_2 \\ t_4 \end{matrix} \right)  \right] \nonumber \\
&=& \frac{\sqrt{\pi T_0}}{m_2} \left(\frac{ m_2^2 T_0}{m_1^2 m_2^2 - \tilde{k}^4}\right)^{m+n+\frac{1}{2}} \Gamma \left(m+n+\frac{1}{2} \right) \;,
\end{eqnarray}
\begin{eqnarray}
\mathcal{\hat{I}}^{(b)}_{t_2} &=& \int_{-\infty}^{\infty} dt_2 \, dt_4 |t_2|^{2m} t_4^{2n} \exp \left[  -\frac{(t_2\; t_4)}{T_0} \left(\begin{matrix} m_1^2 & \tilde{k}^2 \\ \tilde{k}^2 & m_2^2 \end{matrix} \right) \left(\begin{matrix} t_2 \\ t_4 \end{matrix} \right)  \right] \nonumber \\
&=& \frac{T_0^{m+n+1}}{m_1^{2m+1} m_2^{2n+1}} \Gamma \left(m+\frac{1}{2} \right) \Gamma \left(n+\frac{1}{2} \right) \nonumber \\
&& \qquad \times \, _2F_1\left(m+\frac{1}{2},n+\frac{1}{2};\frac{1}{2};\frac{\tilde{k}^4}{m_1^2 m_2^2}\right) \;,
\end{eqnarray}
\begin{eqnarray}
\mathcal{\hat{I}}^{(c)}_{t_2} &=& \int_{-\infty}^{\infty} dt_2 \, dt_4 |t_2|^{2m} t_2 t_4 \exp \left[  -\frac{(t_2\; t_4)}{T_0} \left(\begin{matrix} m_1^2 & \tilde{k}^2 \\ \tilde{k}^2 & m_2^2 \end{matrix} \right) \left(\begin{matrix} t_2 \\ t_4 \end{matrix} \right)  \right] \nonumber \\
&=&  -\frac{\tilde{k}^2 \sqrt{\pi T_0}}{m_2^3} \left(\frac{ m_2^2 T_0}{m_1^2 m_2^2 - \tilde{k}^4}\right)^{m+\frac{3}{2}} \Gamma \left(m+\frac{3}{2} \right) \;,
\end{eqnarray}
and similarly we find,
\begin{eqnarray}
\mathcal{\hat{I}}^{(a)}_{t_4} &=& \int_{-\infty}^{\infty} dt_2 \, dt_4 |t_4|^{2m} t_4^{2n} \exp \left[  -\frac{(t_2\; t_4)}{T_0} \left(\begin{matrix} m_1^2 & \tilde{k}^2 \\ \tilde{k}^2 & m_2^2 \end{matrix} \right) \left(\begin{matrix} t_2 \\ t_4 \end{matrix} \right)  \right] \nonumber \\
&=& \frac{\sqrt{\pi T_0}}{m_1} \left(\frac{m_1^2 T_0}{m_1^2 m_2^2 - \tilde{k}^4}\right)^{m+n+\frac{1}{2}} \Gamma \left(m+n+\frac{1}{2} \right) \;,
\end{eqnarray}
\begin{eqnarray}
\mathcal{\hat{I}}^{(b)}_{t_4} &=& \int_{-\infty}^{\infty} dt_2 \, dt_4 |t_4|^{2m} t_2^{2n} \exp \left[  -\frac{(t_2\; t_4)}{T_0} \left(\begin{matrix} m_1^2 & \tilde{k}^2 \\ \tilde{k}^2 & m_2^2 \end{matrix} \right) \left(\begin{matrix} t_2 \\ t_4 \end{matrix} \right)  \right] \nonumber \\
&=& \frac{T_0^{m+n+1}}{m_1^{2n+1} m_2^{2m+1}} \Gamma \left(m+\frac{1}{2} \right) \Gamma \left(n+\frac{1}{2} \right) \nonumber \\
&& \qquad \times \, _2F_1\left(m+\frac{1}{2},n+\frac{1}{2};\frac{1}{2};\frac{\tilde{k}^4}{m_1^2 m_2^2}\right) \;,
\end{eqnarray}
\begin{eqnarray}
\mathcal{\hat{I}}^{(c)}_{t_4} &=& \int_{-\infty}^{\infty} dt_2 \, dt_4 |t_2|^{2m} t_2 t_4 \exp \left[  -\frac{(t_2\; t_4)}{T_0} \left(\begin{matrix} m_1^2 & \tilde{k}^2 \\ \tilde{k}^2 & m_2^2 \end{matrix} \right) \left(\begin{matrix} t_2 \\ t_4 \end{matrix} \right)  \right] \nonumber \\
&=&  -\frac{\tilde{k}^2 \sqrt{\pi T_0}}{m_1^3} \left(\frac{ m_1^2 T_0}{m_1^2 m_2^2 - \tilde{k}^4}\right)^{m+\frac{3}{2}} \Gamma \left(m+\frac{3}{2} \right) \;.
\end{eqnarray}
Note that we have assumed that $n,m$ are even for all the expressions above, the integrals yield zero otherwise. For two absolute values we can have the integral,
\begin{eqnarray}
\mathcal{\hat{I}}_{t_2, t_4} = \int_{-\infty}^{\infty} dt_2 \, dt_4 |t_2 t_4| \exp \left[  -\frac{(t_2\; t_4)}{T_0} \left(\begin{matrix} m_1^2 & \tilde{k}^2 \\ \tilde{k}^2 & m_2^2 \end{matrix} \right) \left(\begin{matrix} t_2 \\ t_4 \end{matrix} \right)  \right] \;,
\end{eqnarray}
which appears at subsubleading order in \eqref{eq:xtraconti4ltoi4ssl}. We can perform an integral of the form above by expanding,
\begin{equation}
e^{-2 \frac{t_2 t_4}{T_0} \tilde{k}^2} = \sum_{n=0}^{\infty} \frac{1}{(2n)!} \left[ \left( 2 \frac{t_2 t_4}{T_0} \tilde{k}^2 \right)^2 \right]^n + \text{odd terms} \;,
\end{equation}
where we have focused only on even terms as they are the only ones that can contribute to the integral. We therefore have,
\begin{eqnarray}
\mathcal{\hat{I}}_{t_2, t_4} &=& \int_{0}^{\infty} dt_2^2 \, dt_4^2 \sum_{n=0}^{\infty} \frac{1}{(2n)!} \left( 4 \frac{t_2^2 t_4^2}{T_0^2} \tilde{k}^4 \right)^n e^{-m_1^2 \frac{t_2^2}{T_0}} e^{-m_2^2 \frac{t_4^2}{T_0}} \nonumber \\
&=& \int_{0}^{\infty} dx_2 \, dx_4 e^{-x_2} e^{-x_4} \frac{T_0^2}{m_1^2 m_2^2} \sum_{n=0}^{\infty} \frac{x_2^n x_4^n}{(2n)!} \left( 4 \frac{\tilde{k}^4}{m_1^2 m_2^2} \right)^n  \nonumber \\
&=& \frac{T_0^2}{m_1^2 m_2^2} \sum_{n=0}^{\infty} \frac{(n!)^2}{(2n)!} \left( 4 \frac{\tilde{k}^4}{m_1^2 m_2^2}\right)^n \nonumber \\
&=& \frac{T_0^2}{m_1^2 m_2^2 - \tilde{k}^4} + \frac{T_0^2 \tilde{k}^2}{(m_1^2 m_2^2 - \tilde{k}^4)^{3/2}} \arctan \left[\frac{\tilde{k}^2}{\sqrt{m_1^2 m_2^2 - \tilde{k}^4}} \right] \;.
\end{eqnarray}
We also need the following integrals which can be solved in a similar way,
\begin{eqnarray}
\mathcal{\hat{I}}^{(a)}_{t_2, t_4} &=& \int_{-\infty}^{\infty} dt_2 \, dt_4 |t_2 t_4| t_2^2 \exp \left[  -\frac{(t_2\; t_4)}{T_0} \left(\begin{matrix} m_1^2 & \tilde{k}^2 \\ \tilde{k}^2 & m_2^2 \end{matrix} \right) \left(\begin{matrix} t_2 \\ t_4 \end{matrix} \right)  \right] \nonumber \\
&=& \frac{T_0^3 (\tilde{k}^4 +4 m_1^2 m_2^2)}{2 m_1^2 \left(m_1^2 m_2^2-\tilde{k}^4\right)^2} - \frac{3 T_0^3 \tilde{k}^2 m_1^2 m_2^2}{2 m_1^2 \left(m_1^2 m_2^2-\tilde{k}^4\right)^{3/2}} \arctan \left[ \frac{\tilde{k}^2}{\sqrt{m_1^2 m_2^2-\tilde{k}^4}} \right]   \;,
\end{eqnarray}
\begin{eqnarray}
\mathcal{\hat{I}}^{(b)}_{t_2, t_4} &=& \int_{-\infty}^{\infty} dt_2 \, dt_4 |t_2 t_4| t_4^2 \exp \left[  -\frac{(t_2\; t_4)}{T_0} \left(\begin{matrix} m_1^2 & \tilde{k}^2 \\ \tilde{k}^2 & m_2^2 \end{matrix} \right) \left(\begin{matrix} t_2 \\ t_4 \end{matrix} \right)  \right] \nonumber \\
&=& \frac{T_0^3 (\tilde{k}^4 +4 m_1^2 m_2^2)}{2 m_2^2 \left(m_1^2 m_2^2-\tilde{k}^4\right)^2} - \frac{3 T_0^3 \tilde{k}^2 m_1^2 m_2^2}{2 m_2^2 \left(m_1^2 m_2^2-\tilde{k}^4\right)^{3/2}} \arctan \left[ \frac{\tilde{k}^2}{\sqrt{m_1^2 m_2^2-\tilde{k}^4}} \right]   \;,
\end{eqnarray}
\begin{eqnarray}
\mathcal{\hat{I}}^{(c)}_{t_2, t_4} &=& \int_{-\infty}^{\infty} dt_2 \, dt_4 |t_2 t_4| t_2 t_4 \exp \left[  -\frac{(t_2\; t_4)}{T_0} \left(\begin{matrix} m_1^2 & \tilde{k}^2 \\ \tilde{k}^2 & m_2^2 \end{matrix} \right) \left(\begin{matrix} t_2 \\ t_4 \end{matrix} \right)  \right] \nonumber \\
&=& -\frac{3 T_0^3 \tilde{k}^2}{2 \left(m_1^2 m_2^2-\tilde{k}^4\right)^2} - \frac{T_0^3( 2\tilde{k}^2 + m_1^2 m_2^2)}{2\left(m_1^2 m_2^2-\tilde{k}^4\right)^{5/2}} \arctan \left[ \frac{\tilde{k}^2}{\sqrt{m_1^2 m_2^2-\tilde{k}^4}} \right] \;.
\end{eqnarray}

\section{Geodesics in a $D$-dimensional Schwarzschild Background} \label{sec:geodesics}

We can write the action for a massive relativistic particle of mass $m$ in a background $g_{\mu \nu}$ as,
\begin{equation}
S = \frac{1}{2} \int d \tau \left( e(\tau)^{-1} \dot{x}^2 -m^2 e(\tau) \right) \label{eq:geo1} \;,
\end{equation}
where $\dot{x}^2 = g_{\mu \nu} \frac{d x^{\mu}}{d \tau} \frac{d x^{\nu}}{d \tau}$ and $e(\tau)$ is an auxiliary field. The equation of motion for $e(\tau)$ is then easily obtained,
\begin{equation}
e^2 = - \frac{\dot{x}^2}{m^2} \label{eq:geo2} \;.
\end{equation}
Using the equation above we will calculate the deflection angle for a massive probe in the background of a Schwarzschild black hole. The metric for a $D$-dimensional Schwarzschild black hole is given by,
\begin{equation}
g_{\mu \nu} dx^{\mu} dx^{\nu} = - \left(1-\left(\frac{R_s}{r}\right)^n \right) dt^2 + \left(1-\left(\frac{R_s}{r}\right)^n \right)^{-1} dr^2 + r^2 d \Omega_{n+1}^2 \label{eq:geo3} \;,
\end{equation}
where $n=D-3$ and $R_s$ is the Schwarzschild radius.

For simplicity we will use the spherical symmetry of the Schwarzschild solution and work in the equatorial plane and so we set all the angles, $\theta_i= \pi/2$, such that we are left with only one angle in the impact plane $\phi$. Inserting \eqref{eq:geo3} into \eqref{eq:geo2} and using the reparameterization invariance to set $e=1$ we find,
\begin{equation}
-m^2 = - \left(1-\left(\frac{R_s}{r}\right)^n \right) \left( \frac{dt}{d \tau} \right)^2 + \left(1-\left(\frac{R_s}{r}\right)^n \right)^{-1} \left( \frac{dr}{d \tau} \right)^2 + r^2 \left(\frac{d \phi}{d \tau} \right)^2 \label{eq:geo4} \;.
\end{equation}
Since the metric is independent of both $t$ and $\phi$ we have two constants of the motion which are given by,
\begin{eqnarray}
- E &=& - \left(1 - \left(\frac{R_s}{r}\right)^n \right) \;, \frac{dt}{d\tau}  \\
J &=& r^2 \frac{d\phi}{d\tau} \;,
\end{eqnarray}
where $E$ and $J$ parametrize the energy and total angular momentum of the system respectively. Notice that the symbol $E$ in this appendix corresponds to $E_2$ in sections \ref{sec:probeind} and \ref{sec:probeind4} where we consider various probe-limits to the 1PM and 2PM two-body deflection angles. Substituting these quantities into \eqref{eq:geo4} we can find an expression for $dr/d\tau$ and using the chain rule we can find an expression for $d\phi/dr$. The corresponding deflection angle is then given by,
\begin{equation}
\Phi = 2 \int_{r_0}^{\infty} dr \left(\frac{d\phi}{dr} \right) - \pi \;, 
\end{equation}
where $r_0$ is the point of closest approach. Inserting the relevant quantities into the expression above and expanding in powers of $(R_s/r_0)^n$ we notice that we can express the result as a series,
\begin{equation}
\Phi = 2 \sum_{j=1}^{\infty} \int_{0}^{1} du A_j(u) \left(\frac{R_s}{r_0}\right)^{jn} \;,
\end{equation}
where we have explicitly evaluated the integrals up to third order and found the following results,
\begingroup 
\allowdisplaybreaks
\begin{align}
& \int_{0}^{1} du A_1(u) = \frac{\sqrt{\pi} \, \Gamma \left(\frac{n+1}{2}\right) \left(E^2 (n+1)-m^2\right)}{4 \left(E^2-m^2\right) \Gamma \left(\frac{n}{2}+1\right)} \;, \\
\nonumber \\
& \int_{0}^{1} du A_2(u) = \frac{\sqrt{\pi}}{16 \left(E^2-m^2\right)^2} \Biggl(\frac{\Gamma \left(n+\frac{1}{2}\right) \left(E^4 (4 (n+2) n+3)-6 E^2 m^2 (2 n+1)+3 m^4\right)}{\Gamma (n+1)}  \nonumber \\*
& - \frac{4 E^2 \Gamma \left(\frac{n+1}{2}\right) \left(E^2 (n+1)-m^2\right)}{\Gamma \left(\frac{n}{2}\right)}\Biggr)  \;, \\
\nonumber \\
& \int_{0}^{1} du A_3(u) = \frac{\sqrt{\pi }}{32 \left(m^2-E^2\right)^3} \Biggl(-\frac{2 E^2 \Gamma \left(\frac{n+1}{2}\right) \left(E^2 (n-2)+4 m^2\right) \left(E^2 (n+1)-m^2\right)}{\Gamma \left(\frac{n}{2}\right)} \nonumber \\*
& + \frac{2 E^2 \Gamma \left(n+\frac{1}{2}\right) \left(E^4 (4 (n+2) n+3)-6 E^2 m^2 (2 n+1)+3 m^4\right)}{\Gamma (n)} \nonumber \\*
& +\frac{\Gamma \left(\frac{3 n}{2}+\frac{1}{2}\right)}{\Gamma \left(\frac{3 n}{2}+1\right)} \bigl[E^6 (-(n+1)) (3 n+1) (3 n+5)+15 E^4 m^2 (n+1) (3 n+1)\nonumber \\*
& -15 E^2 m^4 (3 n+1)+5 m^6\bigr] \Biggr) \;.
\end{align}
\endgroup
These results in addition to the relation up to third order between the point of closest approach and angular momentum $J$ can be used to express the result for the deflection angle up to third order in $(R_s/J)^n$. The relation between $r_0$ and $J$ is found by evaluating,
\begin{equation}
\frac{dr}{d \tau} \biggr\rvert_{r=r_0} = E^2-m^2 \left(1 - \frac{R_s}{r_0}\right)^n - \frac{J^2}{r_0^2} \left(1 - \frac{R_s}{r_0}\right)^n = 0 
\end{equation}
and up to third order in $J(R_s/J)^n$ is found to be,
\begin{eqnarray}
&& r_0 \approx J \biggl(\frac{1}{\sqrt{E^2-m^2}} - \frac{1}{2} E^2 \left(E^2-m^2\right)^{\frac{n-3}{2}} \left( \frac{R_s}{J} \right)^{n} \nonumber \\
&& - \frac{1}{8} E^2 \left(E^2 (2 n+1)-4 m^2\right) \left(E^2-m^2\right)^{n-\frac{5}{2}} \left(\frac{R_s}{J}\right)^{2 n}\biggr) \;.
\end{eqnarray}
Putting together the relevant quantities and expanding as a power series in $(R_s/J)^n$ we find that the deflection angle is given by,
\begin{eqnarray}
&& \Phi = \sum_{j=1}^{\infty} \Phi_j \left( \frac{R_s}{J} \right)^{j(D-3)} \label{eq:geoPhi} \;,
\end{eqnarray}
where we have substituted for $n=D-3$ and we have for the first 3 terms,
\begin{align}
&\Phi_1 = \frac{\sqrt{\pi } \Gamma \left(\frac{D}{2}-1\right) (E^2-m^2)^{\frac{D-5}{2}} \left[(D-2) E^2-m^2\right]}{2 \Gamma \left(\frac{D-1}{2}\right)} \;, \label{eq:geophi1} \\
\nonumber \\
& \Phi_2 = \frac{\sqrt{\pi } \Gamma \left(D-\frac{5}{2}\right) (E^2-m^2)^{D-5}}{8 \Gamma (D-2)} \nonumber \\
& \times \left[(2 D-5) (2 D-3) E^4+6 (5-2 D) E^2 m^2+3 m^4\right] \;, \label{eq:geophi2} \\
\nonumber \\
& \Phi_3 = \frac{\sqrt{\pi } \Gamma \left(\frac{3 D}{2}-4\right) (E^2-m^2)^{\frac{3 (D-5)}{2}}}{16 \Gamma \left(\frac{3 D}{2}-\frac{7}{2}\right)} \nonumber \\
& \times \bigl[(3 D-8) (3 D-4) (D-2) E^6-15 (D-2) (3 D-8) E^4 m^2 \nonumber \\
& +15 (3 D-8) E^2 m^4-5 m^6\bigr] \label{eq:geophi3} \;.
\end{align}
There are a few limiting cases of the above results that we can use to compare with known results. The result for a null geodesic in general $D$ is given by the $m=0$ case of the equations above. We find,
\begin{eqnarray}
\Phi &=& \frac{\sqrt{\pi } \Gamma \left(\frac{D}{2}\right)}{\Gamma \left(\frac{D-1}{2}\right)}\left( \frac{R_s}{b} \right)^{D-3} + \frac{\sqrt{\pi } \Gamma \left(D-\frac{1}{2}\right)}{2 \Gamma (D-2)} \left( \frac{R_s}{b} \right)^{2(D-3)} \nonumber \\
&&  + \frac{\sqrt{\pi } \Gamma \left(\frac{3 D}{2}-1\right)}{6 \Gamma \left(\frac{3 D}{2}-\frac{7}{2}\right)} \left( \frac{R_s}{b} \right)^{3(D-3)} + \ldots \;, \label{eq:geogendnull}
\end{eqnarray}
where we have also used that when $m=0$ we have $J \simeq Eb$. Comparing the above expression with the results found in appendix D of \cite{Collado:2018isu} we find agreement for the first two terms (note the third term was not calculated in the aforementioned reference). We can also look at the $D=4$ timelike geodesic ($m \neq 0$) case where we find,
\begin{eqnarray}
\Phi &=& \frac{\left(2 E^2-m^2\right)}{\sqrt{E^2-m^2}} \left( \frac{R_s}{J} \right) + \frac{3\pi}{16} \left(5 E^2-m^2\right) \left( \frac{R_s}{J} \right)^2 \nonumber \\
&&  +\frac{\left(64 E^6-120 E^4 m^2+60 E^2 m^4-5 m^6\right)}{12 (E^2-m^2)^{3/2}} \left( \frac{R_s}{J} \right)^3 + \ldots \;, \label{eq:geogendtime4}
\end{eqnarray}
which has been checked and agrees with equivalent results in \cite{Damour:2017zjx,Bern:2019nnu, Antonelli:2019ytb}. Finally by taking the $m=0$ case of the expression above we can also look at the case of a null geodesic in $D=4$. This yields,
\begin{equation}
\Phi = \frac{2R_s}{b} + \frac{15\pi}{16} \left( \frac{R_s}{b} \right)^{2} + \frac{16}{3}\left( \frac{R_s}{b} \right)^{3}  + \ldots  \;. \label{eq:geo4null}
\end{equation}
We find that this agrees with well known results.

\providecommand{\href}[2]{#2}\begingroup\raggedright\endgroup  

\end{document}